\documentclass[journal=gmj]{CUP-JNL-DTM}
\newcommand{\ArtType}{}

\makeatletter
\AtBeginDocument{%
  \let\CUPorigbibliographystyle\bibliographystyle
  \renewcommand{\bibliographystyle}[1]{%
    \CUPorigbibliographystyle{unsrtnat}%
  }%
}
\makeatother
\usepackage{graphicx}
\usepackage{multicol,multirow}
\usepackage{amsmath,amssymb,amsfonts}
\usepackage{mathrsfs}
\usepackage{amsthm}
\usepackage{rotating}
\usepackage{appendix}
\usepackage{ifpdf}
\usepackage[T1]{fontenc}
\usepackage{tabularx}
\usepackage{ragged2e}
\usepackage{newtxtext}
\usepackage{graphicx}
\usepackage{float}
\usepackage{newtxmath}
\usepackage{xurl}
\usepackage{url}
\usepackage{float}
\usepackage{textcomp}
\counterwithout{equation}{chapter}
\usepackage{xcolor}
\usepackage{lipsum}
\usepackage[colorlinks,allcolors=blue]{hyperref}

\theoremstyle{definition}

\numberwithin{equation}{section}

\usepackage{tikz} 
\usetikzlibrary{arrows.meta, positioning}

\renewcommand{\theequation}{\arabic{equation}}

\begin{document}

\begin{Frontmatter}

\title[Article Title]{Knowledge Synthesis Review Framework: Task-Level Benchmarking of LLM-Based Systems for Multi-Source Evidence Synthesis}

\author[1]{Wafa Shafqat}
\author[1,2]{Mark Patterson}
\author[1,2,3,4,5,6]{Steven Liss}


\address[1]{\orgdiv{Magnet}, \orgname{Toronto Metropolitan University}, \orgaddress{\city{Toronto}, \postcode{M5B2K3}, \state{Ontario},  \country{Canada}}}

\address[2]{\orgdiv{Future skill center}, \orgname{Toronto Metropolitan University}, \orgaddress{\city{Toronto}, \postcode{M5B2K3}, \state{Ontario},  \country{Canada}}}

\address[3]{\orgdiv{Office of the Vice-President Research and Innovation}, \orgname{Toronto Metropolitan University}, \orgaddress{\city{Toronto}, \postcode{M5B2K3}, \state{Ontario}, \country{Canada}}}

\address[4]{\orgdiv{Department of Chemistry and Biology}, \orgname{Toronto Metropolitan University}, \orgaddress{\city{Toronto}, \postcode{M5B2K3}, \state{Ontario}, \country{Canada}}}

\address[5]{\orgdiv{Professor Emeritus, Environmental Studies}, \orgname{Queen’s University}, \orgaddress{\city{Kingston}, \postcode{K7L3N6}, \state{Ontario}, \country{Canada}}}

\address[6]{\orgdiv{Professor Extraordinaire, Department of Microbiology}, \orgname{Stellenbosch University}, \country{South Africa}.
\email{steven.liss@torontomu.ca; ovpri@torontomu.ca}}



\abstract{Evidence in rapidly evolving fields is fragmented across academic studies, industry reports, policy documents, and media sources that differ in quality, structure, and purpose, making timely synthesis difficult. Large language models (LLMs) may accelerate this work, but their reliability across the distinct cognitive tasks of a review remains uncertain. We introduce the Knowledge Synthesis Review (KSR), a human-in-the-loop framework that decomposes evidence synthesis into screening, extraction, analysis, and synthesis, benchmarks LLM-based systems on each task against expert reference standards, and routes each task to the best-performing system under continuous expert validation. We evaluated GPT-5, Claude Sonnet 4, Gemini 2.5 Pro, and NotebookLM on a 244-document benchmark subset drawn from a 1,893-document corpus on AI and work spanning four source types, against a gold standard with high inter-rater reliability (92.2\% agreement, $\kappa = 0.80$). No system led on all tasks. Claude Sonnet 4 achieved the highest screening accuracy (82.8\%) and GPT-5 the highest recall (91.8\%) at the expense of lower specificity. Extraction exceeded 90\% agreement for titles and sources but degraded in author and reference fields. Performance declined most in interpretive analysis and cross-source synthesis, where expert judgment remained essential. A contamination check on post-cutoff documents showed no evidence that prior exposure inflated results. Applied to the full corpus, the routed workflow surfaced cross-source asymmetries and blind spots that single-source synthesis would miss, including worker well-being, small firms, and the Global South. KSR offers a transparent, auditable, model-agnostic framework for governing LLM assistance in research synthesis while preserving human accountability.}

\end{Frontmatter}

\section{Introduction}\label{sec1}

Evidence synthesis faces a growing structural problem in fast-moving research domains: the relevant evidence is dispersed across source types that differ fundamentally in quality, structure, incentives, and intended audience. Peer-reviewed research, industry reports, policy briefs, and media commentary each capture part of the picture, yet traditional systematic review methods were designed primarily for relatively homogeneous bodies of academic literature and can take anywhere from six months to over a year to complete \cite{borah2017analysis}. In domains where technologies, organizational practices, and policy debates evolve quickly, this creates a widening lag between evidence generation and evidence synthesis, and it leaves the public conversation to be shaped by whichever individual sources circulate most widely rather than by the balance of evidence.

The relationship between artificial intelligence (AI) and labor markets illustrates this problem sharply, and serves as the demonstration domain for this study. For example, in May 2024, the McKinsey Global Institute projected that generative AI could automate up to 30\% of work hours by 2030 \cite{McKinsey2023GenerativeAI}. A subsequent 2025 study of AI chatbots in Denmark combined survey and administrative data on approximately 25,000 workers across 7,000 workplaces, spanning 11 occupations with high AI exposure \cite{Humlum2025LLMEffects}. It found no statistically significant changes in aggregate output, hours worked, or wages within the first year of adoption. These findings do not contradict McKinsey's long-term automation scenario; they concern a shorter time horizon, a specific set of occupations, and a particular class of tools. However, their reception diverged sharply. The McKinsey projection received broad public and corporate visibility, while the Danish evidence appears to have reached primarily academic and policy audiences. Prior work indicates that media framing shapes public understanding of AI and can influence both adoption and
governance \cite{ouchchy2020ai}. This pattern of uneven reception, where projections tend to reach wider audiences than empirical findings, calls for a more systematic approach to synthesizing evidence across source types, at a pace traditional review methods cannot sustain.

Large language models (LLMs) offer a practical response. They are already embedded in how researchers, analysts, and policymakers process and produce knowledge \cite{khraisha2024can, bolanos2024artificial}, and excluding them from a study of AI-assisted knowledge work would overlook a central feature of how knowledge production is currently evolving. However, using LLMs without a structured framework carries real risks. They can favor fluency over rigor \cite{chelli2024hallucination, ji2023survey}, and blur the boundaries between research
and speculation \cite{ji2023survey, perkins2023academic}. Recent evaluations conclude that LLMs are not yet reliable enough to conduct systematic reviews autonomously \cite{lieberum2025large, clark2025generative}. Moreover, most existing evaluations of LLM assistance in reviews concentrate on individual stages, most often title and abstract screening, within relatively homogeneous and well-structured literatures such as clinical and biomedical research \cite{clark2025generative, delgado2025transforming, li2025enhancing, wang2025accelerating}. Empirical tests of hybrid human-AI review workflows remain limited \cite{clark2025generative, malik2025hybrid}, particularly evaluations that (a) span all core stages of a review, (b) compare multiple LLM-based systems on the same corpus against a common expert-annotated gold standard, and (c) do so over deliberately heterogeneous evidence that mixes peer-reviewed and grey literature. The key methodological challenge, then, is to develop LLM-assisted review processes that are systematic, transparent, comparative, and governed by expert judgment.

\begin{figure}[H]
    \centering
    \includegraphics[width=\linewidth]{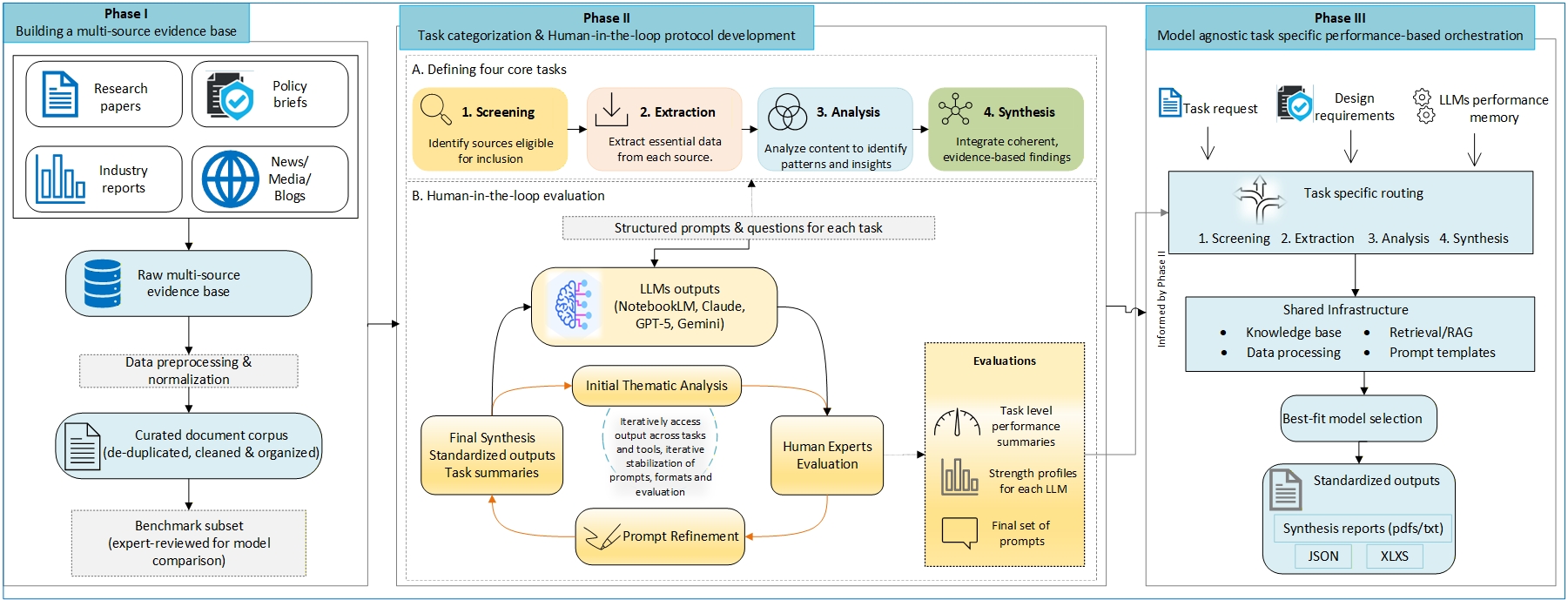}
    \caption{Overview of the three-phase KSR framework. Phase I constructs a multi-source corpus of 1,893 documents and selects 244 for benchmarking. Phase II uses human-in-the-loop evaluation to refine prompts, outputs, and evaluation protocols. Phase III applies model-agnostic, task-specific routing across the corpus.}
    \label{fig:framework_overview}
\end{figure}

To address these gaps, we introduce the knowledge synthesis review (KSR) framework (Fig. \ref{fig:framework_overview}), a structured workflow designed to make LLM-assisted evidence synthesis systematic, replicable, and expert-guided. 
KSR proceeds in three phases. Phase I constructs a multi-source evidence base of 1,893 documents spanning research papers (RPs), industry reports (IRs), policy briefs (PBs), and news, media, and blog sources (NMBs), directly responding to the source fragmentation described above. Phase II decomposes the review process into four cognitive tasks, namely screening, extraction, analysis, and synthesis, and benchmarks four widely used LLM-based systems (GPT-5, Claude Sonnet 4, Gemini 2.5 Pro, and NotebookLM) on each task against gold standards established by expert evaluators on a 244-document benchmark subset.  Phase III operationalizes the benchmark results as a model-agnostic orchestration layer that routes each task to the system with the strongest demonstrated performance for that task, over shared retrieval infrastructure, while maintaining expert validation throughout.

The study addresses two research questions (RQs):
\begin{itemize}
    \item \textbf{RQ1:} How do widely used LLM-based systems perform across the four core tasks of evidence synthesis (screening, extraction, analysis, and synthesis) and across heterogeneous source types, when evaluated against expert-annotated gold standards?
    
    \item \textbf{RQ2:} How can task-level performance differences be translated into a transparent, human-guided orchestration workflow, and what does applying that workflow at scale reveal about the capabilities and limits of LLM-assisted synthesis in a fast-moving, multi-source evidence domain?
\end{itemize}

The paper's primary contribution is methodological: a task-level benchmark of four contemporary LLM-based systems across the core cognitive tasks of evidence synthesis over heterogeneous sources, and, building on it, the KSR framework, a transparent and replicable workflow in which task-specific model routing is governed throughout by expert human validation. Because system capabilities evolve rapidly, KSR is designed to be model-agnostic: its task decomposition, evaluation rubrics, and routing logic can be reapplied as new models appear. As a secondary contribution, we apply KSR end-to-end to the full AI-and-work corpus. This demonstration shows the framework operating on real, uneven evidence and reveals how different source types emphasize different aspects of the same phenomenon, a pattern that synthesis from any single source type would miss. We therefore present the resulting labor-market observations as outputs of the evidence synthesis, useful for identifying patterns, asymmetries, and gaps across sources, rather than as direct empirical estimates of AI's labor-market effects.

\section{Methods}\label{sec2}

The proposed KSR framework was implemented in three sequential phases as illustrated in Fig.~\ref{fig:framework_overview}. Phase I focused on corpus construction, in which a multi-source evidence base was developed, cleaned, and organized into a curated document corpus, from which a benchmark subset was selected for expert review and model comparison.
Phase II focused on task categorization and human-in-the-loop protocol development. In this phase, we first categorize the review process into four core knowledge synthesis tasks. Each task was conducted by expert evaluators and LLMs using 
structured prompts, standardized output formats, and iterative refinements based on model evaluations. Phase III addressed the model-agnostic task orchestration and scaling. The results of phase II informed the task-specific routing across the curated corpus. Rather than relying on a single model for all review stages, the framework assigned each task to the model that demonstrated the strongest empirical performance while preserving human oversight throughout\footnote{The use of LLMs in this study was part of the research methodology and is described throughout this section. All LLM-assisted outputs were subject to human review, validation, and author accountability.}.

\subsection{Phase I: Knowledge Base Creation}

We constructed a multi-source corpus to capture diverse perspectives on the evolving relationship between AI and labor markets. Overall, the full corpus comprises 1,893 documents spanning RPs, IRs, PBs, and NMBs (2020–2025). Table \ref{tab1} shows the composition of the entire corpus from which the benchmark subset was selected for human evaluation and model comparison. The distribution reflects the actual landscape of publicly available evidence on AI and labor markets, where news and media sources substantially outnumber academic and policy documents. We preserved this imbalance to represent the evidence environment as it exists. The smaller samples of IRs and PBs (n=30 each) reflect the limited volume of such documents publicly available on this topic rather than a sampling choice.

For academic research, we selected sources from leading databases, including Web of Science, ScienceDirect, Google
Scholar, IEEE Xplore, Nature, Science, and Springer. We prioritized Q1–Q2 journals, to ensure
high research quality, with the exception of a few Q3 articles, for their citation strength or topical
novelty. Industry reports were chosen for their authority and relevance from organizations such
as National Bureau of Economic Research\footnote{https://www.nber.org/}, McKinsey \& Company\footnote{https://www.mckinsey.com}, Bureau of Labor Statistics\footnote{https://www.bls.gov/}, World Economic Forum\footnote{https://www.weforum.org/}, International Labour Organization (ILO)\footnote{https://www.ilo.org/}, PricewaterhouseCoopers
(PwC)\footnote{https://www.pwc.com/ca/en.html}, LinkedIn\footnote{https://www.linkedin.com/}, Organization for Economic Co-operation and Development (OECD)\footnote{https://www.oecd.org/}
, Deloitte\footnote{https://www.deloitte.com/}, The International Monetary Fund\footnote{https://www.imf.org/en/home}. Policy briefs were collected from Economic
Policy Institute\footnote{https://www.epi.org/}, Massachusetts Institute of Technology (MIT)\footnote{https://web.mit.edu/}, Stanford Human-Centered
Artificial Intelligence (HAI)\footnote{https://hai.stanford.edu/}, OECD, ILO. Media and commentary included publications from
Forbes\footnote{https://www.forbes.com/}, Brookings\footnote{https://www.brookings.edu/}, International Business Machines Corporation (IBM)\footnote{https://www.ibm.com/}, Google News\footnote{https://news.google.com/},
and Chicago Booth Review\footnote{https://www.chicagobooth.edu/review}.

\subsection{Phase II: Task Categorization and Human-in-the-Loop Protocol Development} \label{sec:hl}

\subsubsection{Human-in-the-Loop Benchmark Design}
For human-in-the-loop evaluation, we assembled a 244-document benchmark subset. It includes 103 research papers (RP), 30 industry reports (IR), 30 policy briefs (PB), and 81
news/media/blogs (NMB) articles, as shown in Table \ref{tab1}. Documents were identified via systematic keyword searches (AI and labor market, impacts of AI on work, societal impacts of AI, AI and the future of work). This curated dataset enables a balanced view across scholarly, industrial, policy, and public discourse. Because industry reports and policy briefs were scarce, all available documents of these two types were included; research papers and news/media sources were subsampled to keep expert review tractable while retaining sufficient documents per source type for meaningful cross-model comparison. The benchmark subset therefore deliberately over-represents the scarce, higher-credibility source types relative to the full corpus, in order to ensure each source type was adequately evaluated.

\begin{table}[h]
\centering
\caption{Knowledge types, selection criteria, and document counts before and after processing}
\label{tab1}

\footnotesize
\setlength{\tabcolsep}{3pt}
\renewcommand{\arraystretch}{1.15}

\begin{tabularx}{\linewidth}{
@{}
>{\raggedright\arraybackslash}p{2.6cm}
X
>{\centering\arraybackslash}p{1.5cm}
>{\centering\arraybackslash}p{1.8cm}
>{\centering\arraybackslash}p{1.6cm}
@{}
}
\hline

\textbf{Source Type} &
\textbf{Predefined criteria} &
\textbf{Full Corpus (n)} &
\textbf{Benchmark subset} \\

\hline

RP (Research papers) &
Peer-reviewed, methodological transparency, academic venue &
324 & 103  \\

IR (Industry reports) &
Firm-produced, forward-looking, profit-linked claims &
30 & 30  \\

PB (Policy briefs) &
Government or NGO-affiliated, policy-oriented, implementation-focused &
30 & 30  \\

NMB (News, media, and blogs) &
Non-peer-reviewed, narrative-driven, public engagement goal &
1509 & 81  \\

\hline

\textbf{Total} &
&
\textbf{1893} &
\textbf{244} \\

\hline
\end{tabularx}
\end{table}

All tasks in the KSR workflow involve interaction between LLM-generated outputs and human review checkpoints. The framework was designed to incorporate structured human oversight, in which expert reviewers assess model outputs at each stage of the workflow using predefined task-specific criteria. This human-in-the-loop design ensures that model performance is not evaluated only by whether an output is fluent or complete, but also by whether it is accurate, relevant, source-grounded, and useful for knowledge synthesis. Rather than reserving human evaluation for the final synthesis stage, oversight was embedded throughout to prevent error propagation, as inaccuracies in early stages such as screening and extraction can systematically distort downstream analysis and synthesis.

\begin{figure}[]
    \centering

    \begin{minipage}{0.75\linewidth}
        \centering
        \includegraphics[width=\linewidth]{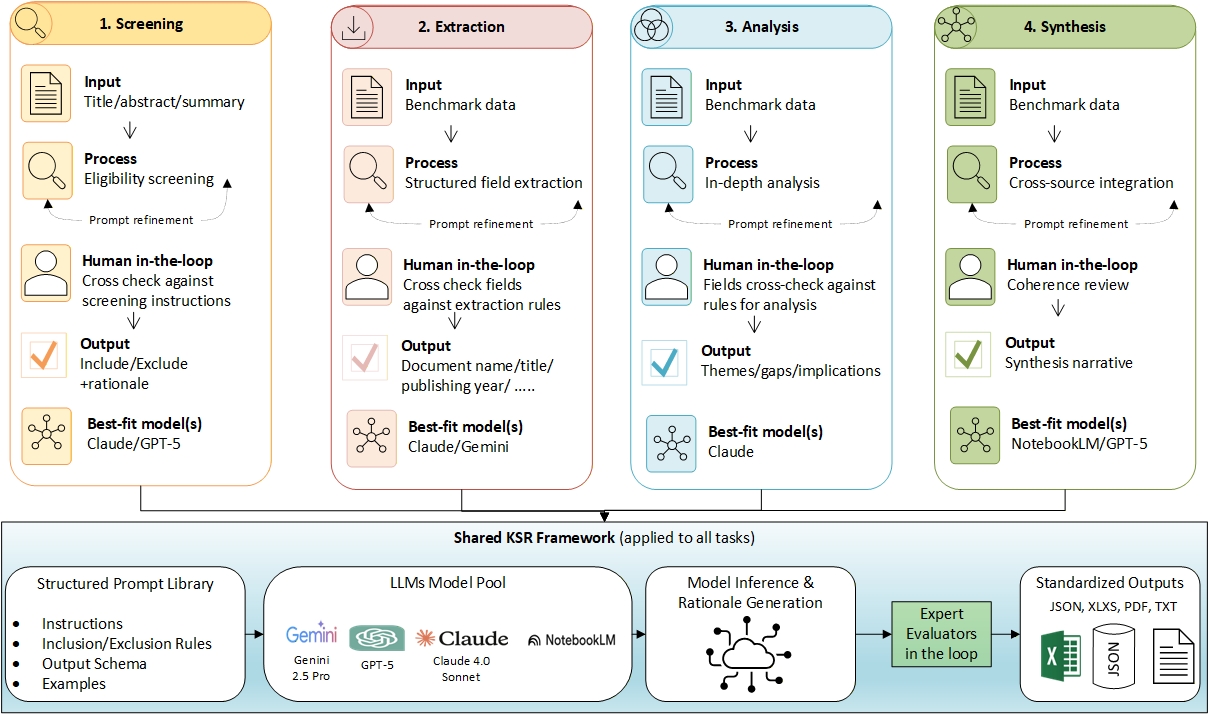}
        \vspace{2pt}
        \textbf{(a)}
    \end{minipage}

    \vspace{8pt}

    \begin{minipage}{0.75\linewidth}
        \centering
        \includegraphics[width=\linewidth]{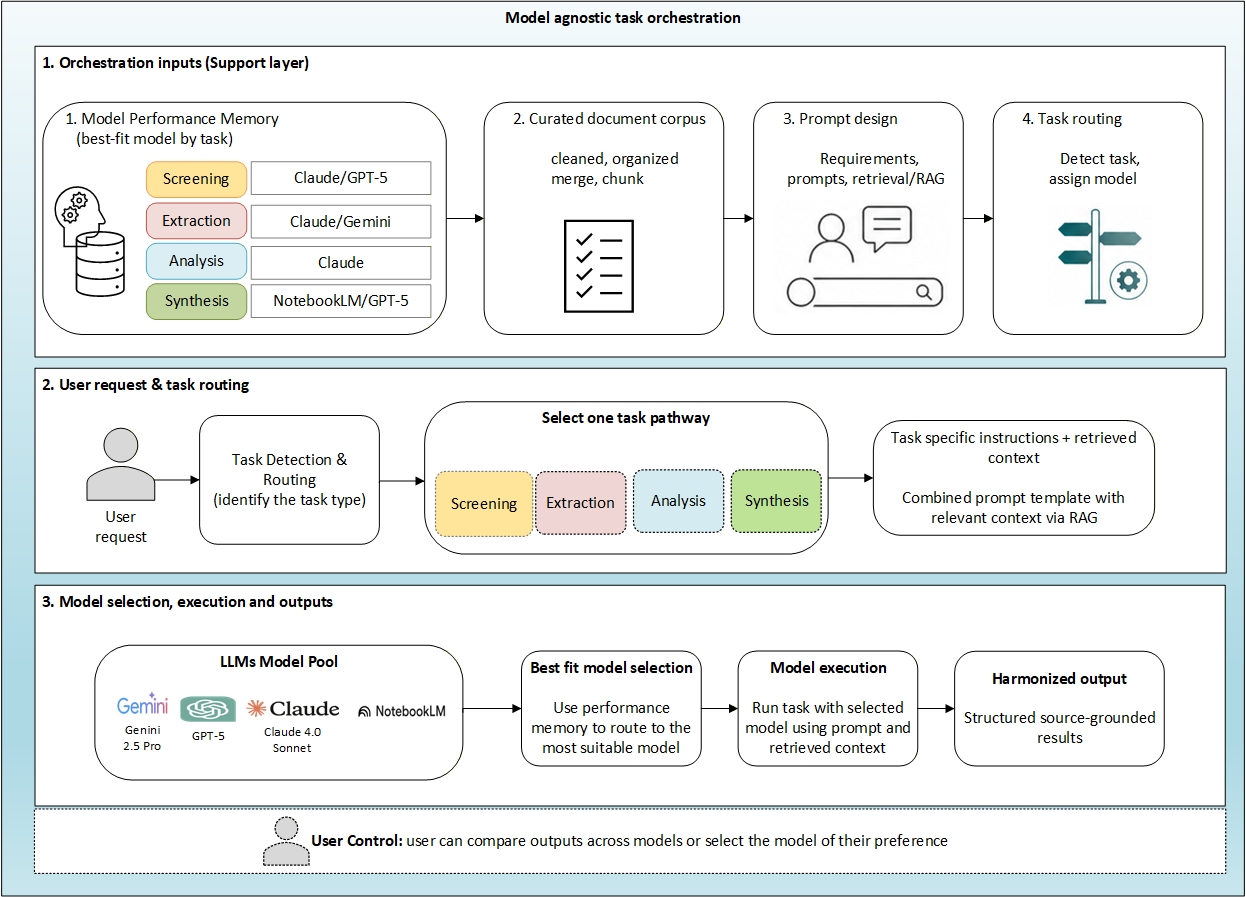}
        \vspace{2pt}
        \textbf{(b)}
    \end{minipage}
    \caption{The KSR framework: (a) Four LLMs perform screening, extraction, analysis, and synthesis on the same sources using standardized prompts, with expert refinement and evaluation producing task-level performance profiles. (b) These results guide task-specific model routing and generate harmonized outputs in formats such as XLSX and JSON}
    \label{fig:fig-2}
\end{figure}

Primary benchmark evaluation was conducted by two field experts, one member of the author team and one evaluator external to, and independent of, the project with expertise in the AI field. To limit bias toward any particular system, all model outputs were anonymized and randomized prior to evaluation, so that evaluators could not identify which system had produced a given output. To assess the reliability of the screening gold standard, the two expert reviewers independently labeled each screened document as include or exclude prior to any reconciliation. Inter-rater reliability was measured using Cohen's kappa, calculated for each source type and across all screened documents. We also report raw percent agreement, because kappa is sensitive to skewed label distributions. Across 244 screened documents, the two reviewers agreed on 92.2 percent of decisions (kappa = 0.804). Agreement was stable across source types, with kappa values of 0.718 for policy briefs, 0.831 for industry reports, 0.818 for research papers, and 0.812 for news and media sources, indicating that the screening criteria were applied consistently across heterogeneous evidence types. Documents on which the reviewers initially disagreed were resolved through consensus to produce the final screening decisions; 183 documents (RP=79, IR=21, PB=21, NMB=62) were retained for extraction, analysis, and synthesis after deduplication.
For the extraction task, model outputs were scored against the source documents by comparing each extracted field to the verified record, with matches graded as exact, partial, or absent. For the analysis and synthesis tasks, outputs were assessed by the two reviewers against predefined rubrics, and scores were reconciled through the same consensus process. Each task was evaluated using structured prompts, standardized output formats, and predefined scoring rubrics. Depending on the task, outputs were assessed on dimensions such as accuracy, completeness, clarity, relevance, source fidelity, reasoning quality, and usefulness. Quantitative ratings, including one-to-five scores where appropriate, were combined with qualitative reviewer comments to identify task-level strengths, weaknesses, and recurring failure patterns across models. Each task in the KSR workflow (Fig. \ref{fig:fig-2}) has distinct evaluation criteria. Screening evaluates whether sources are correctly included or excluded according to predefined eligibility criteria and whether the rationale is appropriate.

Extraction assesses the accuracy and completeness of structured fields, including metadata, methods, findings, and links to supporting evidence. Analysis focuses on interpretive quality, thematic relevance, methodological insight, and the ability to identify limitations, gaps, and implications. Synthesis evaluates coherence, coverage, integration across sources, reduction of redundancy, and the quality of broader conclusions drawn from multiple evidence streams.

\subsubsection{Task-Specific Review Protocols}

\textbf{Task 1: Screening}
Screening was conducted to determine whether each document should be included in the KSR corpus for further extraction, analysis, and synthesis. Each document was assigned a binary label of include or exclude. RPs were screened using titles, abstracts, and keywords, while PBs, IRs, and NMBs were screened using titles and executive summaries. To standardize model inputs, these fields were converted into JSON files and provided to each LLM using the same screening prompt structure.

The screening prompts reflected the manual inclusion and exclusion criteria used by the expert reviewers. Documents were required to be written in English and to demonstrate a clear theoretical, empirical, or policy connection between AI and labor market outcomes. Documents were included if they substantively addressed topics such as automation, employment, wages, skills, productivity, occupational change, organizational transformation, inequality, reskilling, or labor-market policy. Documents were excluded when the available title, abstract, keywords, or summary did not establish a substantive connection to AI-related labor-market outcomes.

The screening protocol was refined iteratively during the benchmark process. Early versions of the prompts included article links, but this approach was unreliable because most models could not consistently access or interpret external sources, with the exception of GPT-5 with web search enabled. In some cases, models generated judgments based only on titles rather than the full screening fields available in the JSON input (Appendix Fig. A1-A3). Additional issues included overlapping outputs, inconsistent rationales, and formatting errors. To address these limitations, the final protocol used simplified natural-language instructions, JSON-formatted inputs, and standardized .xlsx outputs. This stepwise protocol improved consistency across models and allowed screening decisions, rationales, and model outputs to be systematically compared (Appendix Fig. A4). 

\textbf{Task 2: Extraction}
The extraction task evaluated the ability of LLMs to convert unstructured or semi-structured source documents into standardized metadata records. Extraction prompts were tailored to each document type to reflect differences in source structure and available information (Appendix Fig. A5). For RPs, models extracted the document name, title, author(s), publication source, year, and the number of tables, figures, and references. For IRs and PBs, extraction focused on the document name, title, author(s), issuing organization, and year. 

For NMBs, the schema captured the document name, title, issuing body or publisher, publication date, authorship, and links to relevant scholarly sources where available.
The extracted fields were returned in structured formats, such as JSON or spreadsheet-compatible outputs, to support downstream analysis and comparison across models. Model outputs were evaluated against the source documents to assess field accuracy, completeness, consistency with the required schema, and source reliability. This task, therefore, tested not only whether models could identify relevant metadata but also whether they could produce standardized, machine-readable records suitable for integration into the broader KSR workflow.

\textbf{Task 3: Analysis}
The analysis task evaluated the ability of LLMs to identify and summarize substantive, interpretive content from each source type (Appendix Fig. A6). Unlike the extraction task, which focused on structured metadata, the analysis task required models to interpret document content and identify source-specific analytical elements relevant to AI and labor market impacts.
For RPs, models extracted three key dimensions: study limitations, future research directions, and a summary of the methodology. For IRs, the analysis focused on current limitations, key findings, and recommendations. For PBs, models identified gaps in existing policies, proposed interventions, and anticipated outcomes. For NMBs, analysis focused on reported limitations, core findings, and future implications.
All outputs were returned as structured JSON objects to support comparison across models and integration into the synthesis stage. This task tested whether LLMs could move beyond factual extraction to produce structured analytical summaries that preserved the meaning and context of the original sources.

\textbf{Task 4: Synthesis}
The synthesis task evaluated the ability of LLMs to produce integrative, cross-source narratives from the curated document corpus (Appendix Fig. A7). Unlike the analysis task, which focused on source-specific analytical elements, synthesis required models to integrate evidence across RPs, IRs, PBs, and NMBs.
Synthesis prompts guided the models to address six core dimensions: overarching labor-market impacts of AI, stakeholder narratives, cross-source integration, temporal patterns, governance and ethics, and key challenges. Models were also asked to identify proposed future directions and blind spots in the evidence base. Outputs were returned as structured narratives to support comparison across models and integration into the final KSR findings.
The synthesis task used the preprocessed and chunked corpus described in Section 2.1, allowing models to retrieve and integrate relevant evidence across the full document set.
The synthesis outputs were evaluated for coherence, coverage, source integration, reduction of redundancy, source reliability, and usefulness for generating higher-level conclusions. This task tested whether LLMs could move beyond document-level analysis to produce integrated knowledge claims across heterogeneous evidence streams.

\subsubsection{Model Selection and Comparative Evaluation}
We evaluated four widely used LLM-based systems including: GPT-5, Claude Sonnet 4,
Gemini 2.5 Pro, and NotebookLM. These models were selected to represent the leading general-purpose LLMs available at the time of the study, based on their
widespread use and distinct functional profiles. Each model was queried via API or web
interface using standardized and identical prompts. To reduce evaluator bias, outputs were anonymized and randomized before assessment. This design provided a controlled basis for comparing model performance across the four KSR tasks.

\noindent

\subsubsection{Evaluation Metrics and Statistical Analysis}

Performance was assessed using task-specific evaluation metrics. Performance in the screening stage was evaluated using a confusion matrix based on the expert-labeled gold standard. A document was considered a True Positive (TP) when both the LLM and the expert classified it as Include. A False Positive (FP) occurred when the LLM classified a document as Include but the expert classified it as Exclude. A False Negative (FN) occurred when the LLM classified a document as Exclude but the expert classified it as Include. Finally, a True Negative (TN) occurred when both the LLM and the expert classified a document as Exclude. From these quantities, we computed the following evaluation metrics:

\begin{itemize}
    \item Precision: To measure the proportion of documents predicted as relevant that were actually relevant. A higher precision indicates greater resistance to false positives.
    \begin{equation}
    \label{eq:1}
    \mathrm{Precision}=\frac{TP}{TP+FP}
    \end{equation}

    \item Recall: To measure the proportion of truly relevant documents that were successfully identified by the model. A higher recall indicates a lower risk of missing relevant evidence.
    \begin{equation}
    \mathrm{Recall}=\frac{TP}{TP+FN}
    \end{equation}

    \item Specificity: To evaluate the proportion of irrelevant documents that were correctly excluded.
    \begin{equation}
    \mathrm{Specificity}=\frac{TN}{TN+FP}
    \end{equation}

    \item Accuracy measures the overall proportion of correctly classified documents, considering both relevant and irrelevant studies.
    \begin{equation}
    \mathrm{Accuracy}=\frac{TP+TN}{TP+TN+FP+FN}
    \end{equation}

    \item F1-score is the harmonic mean of precision and recall, providing a balanced measure when both false positives and false negatives are important.
    \begin{equation}
    F_1 = \frac{2 \times \mathrm{Precision} \times \mathrm{Recall}}
    {\mathrm{Precision}+\mathrm{Recall}}
    \end{equation}
\end{itemize}

For extraction, we measured
accuracy at the field level. For fixed metadata fields, such as author names, publication years, dates, and document titles, a correct extraction required an exact string match with the expert-verified record. For open-ended or variable text fields, matches were graded as PERFECT, SEMI, or NONE, reflecting full, partial, or absent correspondence with the target information.

For analysis, the model output was evaluated for each document, field, and LLM. The expert reviewers assessed the quality of the generated response to determine whether an output was relevant, sufficiently detailed, well developed, credible, clearly written, and appropriate for the type of source and analytical field being evaluated.
Each output was assessed using six dimensions: relevance, specificity, completeness, evidence, clarity \& structure, and a field-specific criterion.

Each dimension was scored on a five-point scale. A score of 1 indicated that the expected content was largely absent, incorrect, or off-topic. A score of 2 indicated a limited attempt with substantial weaknesses. A score of 3 represented an acceptable response that was relevant but remained generic, incomplete, or uneven. A score of 4 indicated a strong response that addressed most expectations but retained minor gaps. A score of 5 represented an excellent response that was highly relevant, specific, comprehensive, credible, clearly structured, and appropriate for the assigned source category and field. Responses containing no substantive content, or fewer than ten words, received a score of 1 across all dimensions because there was insufficient information for meaningful evaluation. The overall score was calculated as the mean of the six evaluation dimensions and rounded to the nearest whole number.

For synthesis, the outputs were evaluated according to their ability to integrate evidence across source types and generate coherent higher-level conclusions. The evaluation criteria included balance between critique and technical detail, quality of source integration, coverage of key themes, narrative coherence, reduction of redundancy, and usefulness to identify broader patterns, gaps, and future directions.

\subsubsection{Validation and Bias Mitigation}

To strengthen methodological integrity, the KSR framework adopted a multi-stage validation and bias mitigation process. First, all outputs generated by LLMs were reviewed against predefined task-specific criteria. Screening decisions, extracted data, analytical summaries, and synthesis results were evaluated for accuracy, completeness, relevance, source reliability, and interpretive quality. Second, conflicting findings were flagged and retained rather than merged into a single narrative. For example, claims about productivity gains were considered alongside evidence on wage stagnation, job insecurity, occupational polarization, and unequal benefit distribution. This method allowed the synthesis to preserve tensions across different source types, stakeholder viewpoints, and time frames. Third, source credibility was assessed based on criteria like peer-review status, transparency of methods, institutional affiliation, and potential conflicts of interest. This was important because the corpus included diverse evidence from academic research, policy briefs, industry reports, and news/media/blog sources. The analytic rubric enabled consistent evaluation of interpretive quality. Synthesis prompts integrated evidence across different time frames while maintaining contradictions and gaps. Task-specific model routing further minimized reliance on any single model, reducing the impact of model-specific weaknesses on the overall review. The validated process was first tested on the 244-document benchmarking corpus and then applied to the full corpus. This provided a controlled basis for comparing model performance and for generating the main synthesis reported in the results.

A recognized risk when evaluating commercial LLMs on publicly available documents is training data contamination: because the corpus consists of public documents published between 2020 and 2025, some may have been part of the models' training data, so measured extraction performance could reflect prior exposure rather than genuine processing of the supplied text. To assess this, we compared extraction accuracy on documents published after the models' training cutoffs, which they could not have encountered during training, against accuracy on the full corpus. We took the most recent training cutoff among the evaluated systems (April 2025) as the threshold, classifying documents published after it as post-cutoff. The three systems with defined training cutoffs (GPT-5, Claude Sonnet 4, Gemini 2.5 Pro) were assessed; NotebookLM was excluded because it performs its own retrieval over uploaded documents and does not expose a comparable training cutoff. The check was conducted on news and media sources, which had enough post-cutoff documents to support a stable comparison; research papers, policy briefs and industry reports had too few documents published after the cutoff to assess separately. Post-cutoff documents were identified by verified original publication date. Extraction accuracy was computed field by field on identical prompts and inputs, differing only in publication date, and pooled per system for comparison (see Table \ref{tab:model_performance}).

\begin{table}[h]
\centering
\caption{Field-level extraction accuracy for NMB documents on the full corpus and pre- and post-cutoff subsets, by model}
\label{tab:model_performance}

\footnotesize
\setlength{\tabcolsep}{6pt}
\renewcommand{\arraystretch}{1.15}

\begin{tabular}{@{}llcccc@{}}
\toprule

\textbf{Model} &
\textbf{Field} &
\textbf{Full (\%)} &
\textbf{Pre (\%)} &
\textbf{Post (\%)} &
\textbf{Post N} \\

\midrule

\multirow{6}{*}{GPT-5}
& Title  & 98.4 & 97.1 & 100.0 & 27 \\
& Year   & 67.7 & 57.1 & 81.5  & 27 \\
& Source & 96.8 & 94.3 & 100.0 & 27 \\
& Author & 62.9 & 48.6 & 81.5  & 27 \\
& Links  & 74.2 & 77.1 & 70.4  & 27 \\
& \textbf{Total} & \textbf{80.0} & \textbf{74.9} &
\textbf{86.7} & \textbf{135} \\

\midrule

\multirow{6}{*}{Claude}
& Title  & 95.2 & 91.4 & 100.0 & 27 \\
& Year   & 71.0 & 57.1 & 88.9  & 27 \\
& Source & 91.9 & 85.7 & 100.0 & 27 \\
& Author & 67.7 & 54.3 & 85.2  & 27 \\
& Links  & 75.8 & 80.0 & 70.4  & 27 \\
& \textbf{Total} & \textbf{80.3} & \textbf{73.7} &
\textbf{88.9} & \textbf{135} \\

\midrule

\multirow{6}{*}{Gemini}
& Title  & 96.8 & 94.3 & 100.0 & 27 \\
& Year   & 74.2 & 60.0 & 92.6  & 27 \\
& Source & 95.2 & 94.3 & 96.3  & 27 \\
& Author & 66.1 & 51.4 & 85.2  & 27 \\
& Links  & 40.3 & 31.4 & 51.9  & 27 \\
& \textbf{Total} & \textbf{74.5} & \textbf{66.3} &
\textbf{85.2} & \textbf{135} \\

\bottomrule
\end{tabular}

\end{table}

\subsection{Phase III: Model-Agnostic Task Orchestration and Scaling}

Phase III applied the validated KSR workflow from the benchmark subset to the full corpus using a model-agnostic orchestration strategy. This phase used the finalized task-specific prompts, preprocessed document chunks, and model performance summaries generated during Phase II to guide task-specific model routing.
A central component of this phase was a shared retrieval-augmented generation (RAG) infrastructure. The curated corpus was cleaned, semantically chunked, validated, de-duplicated, and stored in ChromaDB\footnote{https://www.trychroma.com/products/chromadb}. This retrieval layer was shared across models, meaning that GPT-5, Claude Sonnet 4, Gemini 2.5 Pro, and NotebookLM were provided access to the same underlying evidence base. The purpose of this shared retrieval layer was to keep model comparison and task execution grounded in a common corpus, reduce dependence on each model’s internal knowledge, and minimize variation caused by unequal access to evidence.
The model performance summaries served as performance memory, capturing which models performed strongest for screening, extraction, analysis, and synthesis. Screening tasks were routed based on classification performance, including precision, recall, accuracy, and specificity. Extraction tasks were routed based on field-level accuracy and schema consistency. Analysis tasks were routed based on relevance, completeness, source reliability, interpretive quality, and clarity. Synthesis tasks were routed based on coherence, source integration, thematic coverage, and the ability to preserve contradictions and blind spots.

For each routed task, the orchestration layer combined three elements: the task-specific prompt template, the relevant retrieved context from the shared RAG layer, and the best-fit model identified through performance memory. Outputs were then returned in harmonized formats, including .xlsx screening results, structured JSON extraction and analysis outputs, and narrative synthesis files. Human oversight remained part of the workflow, with expert review used to validate routed outputs, monitor recurring failure modes, and ensure alignment with the task criteria established in Phase II. 

The shared retrieval layer was used by the three systems queried via API or standard interface (GPT-5, Claude Sonnet 4, Gemini 2.5 Pro). NotebookLM, which performs its own internal retrieval over uploaded documents and exposes no interface to an external vector store, was instead given the same source documents and operated over its native retrieval. Rather than relying on one model across all review tasks, KSR uses shared retrieval infrastructure and task-specific routing to combine model strengths while preserving source grounding, comparability, and human review.

\section{Results}
This section presents the findings organized around the two research questions 
introduced above. We first address RQ1 by reporting model-level benchmark 
performance across the four KSR tasks. We then address 
RQ2 by presenting the human-guided synthesis of labor-market evidence across six 
analytical sub-questions.

\subsection{Model Performance Varies Systematically by Task and Source Type}
\label{sec:rq1}

\begin{figure}[]
    \centering
    \begin{minipage}[t]{0.48\textwidth}
        \centering
        \includegraphics[width=\linewidth]{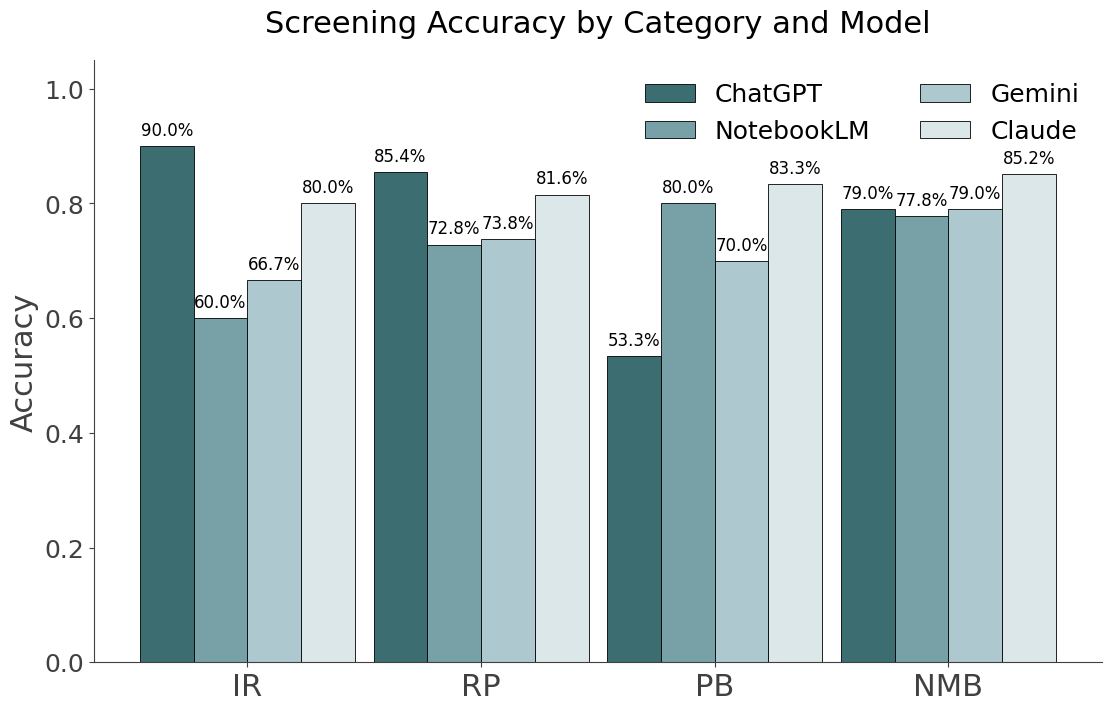}
        \vspace{2pt}
        \textbf{(a)}
        \label{fig:3a}
    \end{minipage}
    \hfill
    \begin{minipage}[t]{0.48\textwidth}
        \centering
        \includegraphics[width=\linewidth]{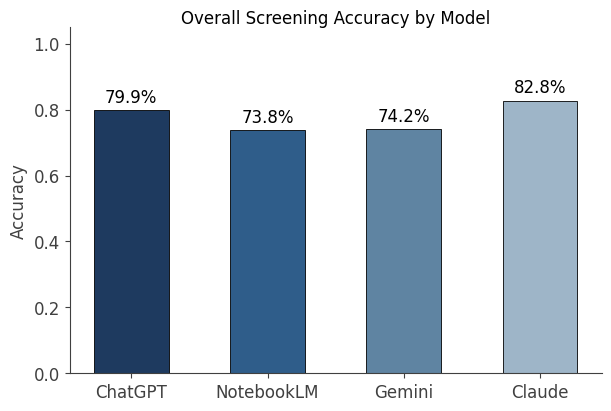}
        \vspace{2pt}
        \textbf{(b)}
        \label{fig:3b}
    \end{minipage}
    \caption{Screening performance evaluation: (a) Accuracy of GPT-5, Claude, 
    Gemini, and NotebookLM across four document categories: IR, NMB, PB, and RP. 
    (b) Overall accuracy of LLMs averaged across all datasets. Claude achieved 
    the highest overall accuracy (82.8\%), followed by GPT-5 (ChatGPT) (Claude Sonnet 4.9\%), 
    Gemini (74.2\%), and NotebookLM (73.8\%)}
    \label{fig:3}
\end{figure}

\begin{figure*}[t]
    \centering
    \includegraphics[width=0.9\textwidth]{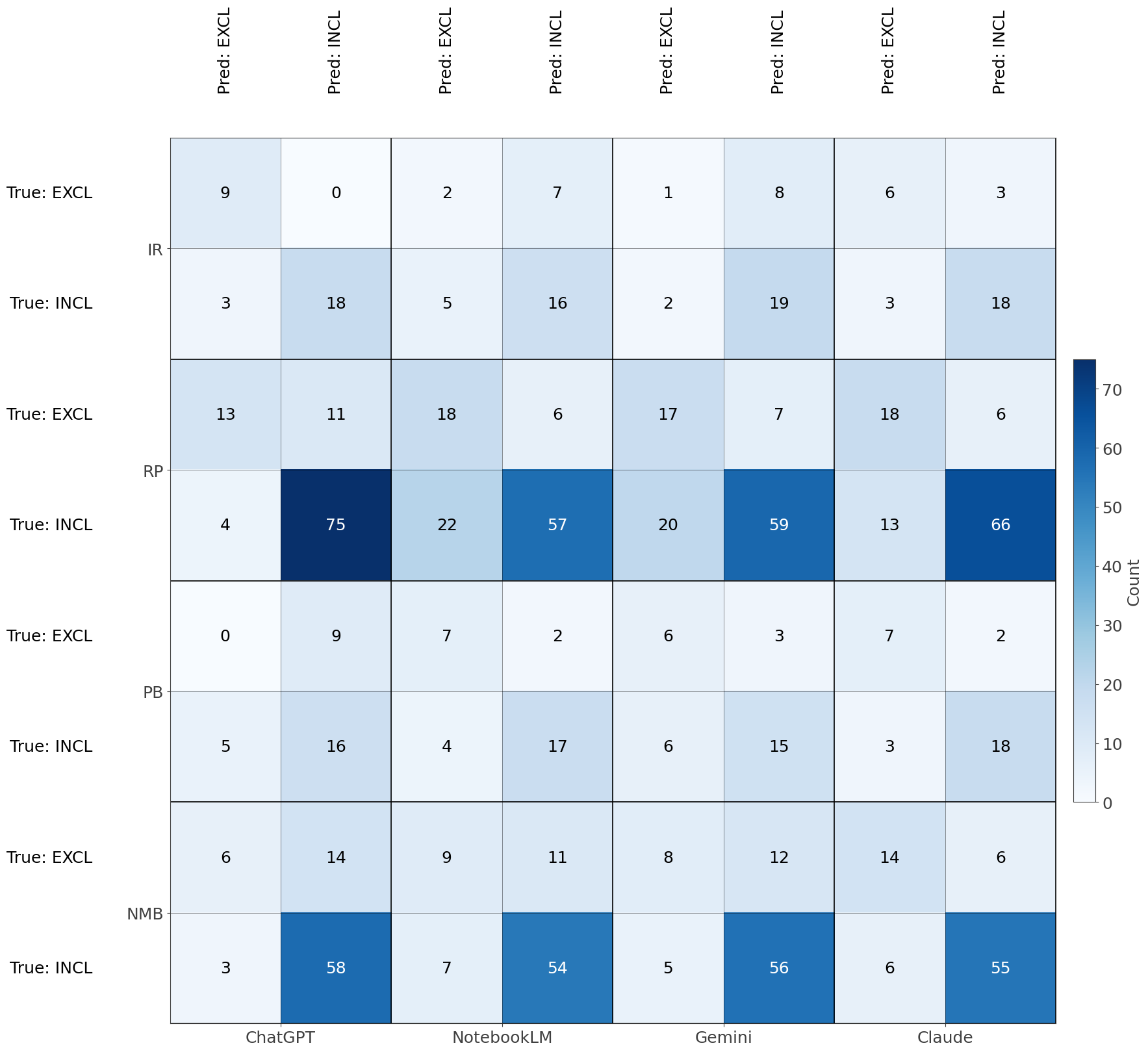}
    \caption{Confusion matrices for screening performance by dataset and model. 
    Each matrix displays true vs.\ predicted inclusion (INCL) and exclusion 
    (EXCL) decisions across four document categories. Across datasets, all models 
    show stronger recall for ``INCL'' decisions than for ``EXCL'' decisions}
    \label{fig:4}
\end{figure*}

In screening, Claude Sonnet 4 achieved the most balanced performance, with an accuracy of 82.8\% (Fig. \ref{fig:3}) and the most balanced performance across metrics. It consistently constrained false positives, particularly in PB and NMB, where precision reached 70\%, making it well-suited to final filtering stages where over-inclusion is costly. GPT-5, in contrast, demonstrated the strongest recall (91.8\%) and excelled in RP and IR, but its specificity collapsed in PB (53.3\%), where it misclassified all true exclusions as inclusions. Gemini 2.5 Pro and NotebookLM adopted more conservative strategies, reducing false positives at the cost of higher false negatives (Fig. \ref{fig:4}). Together, these patterns confirm that screening behavior is context dependent.

\begin{figure*}[t]
    \centering
    \begin{minipage}{0.32\textwidth}
        \centering
        \includegraphics[width=\linewidth]{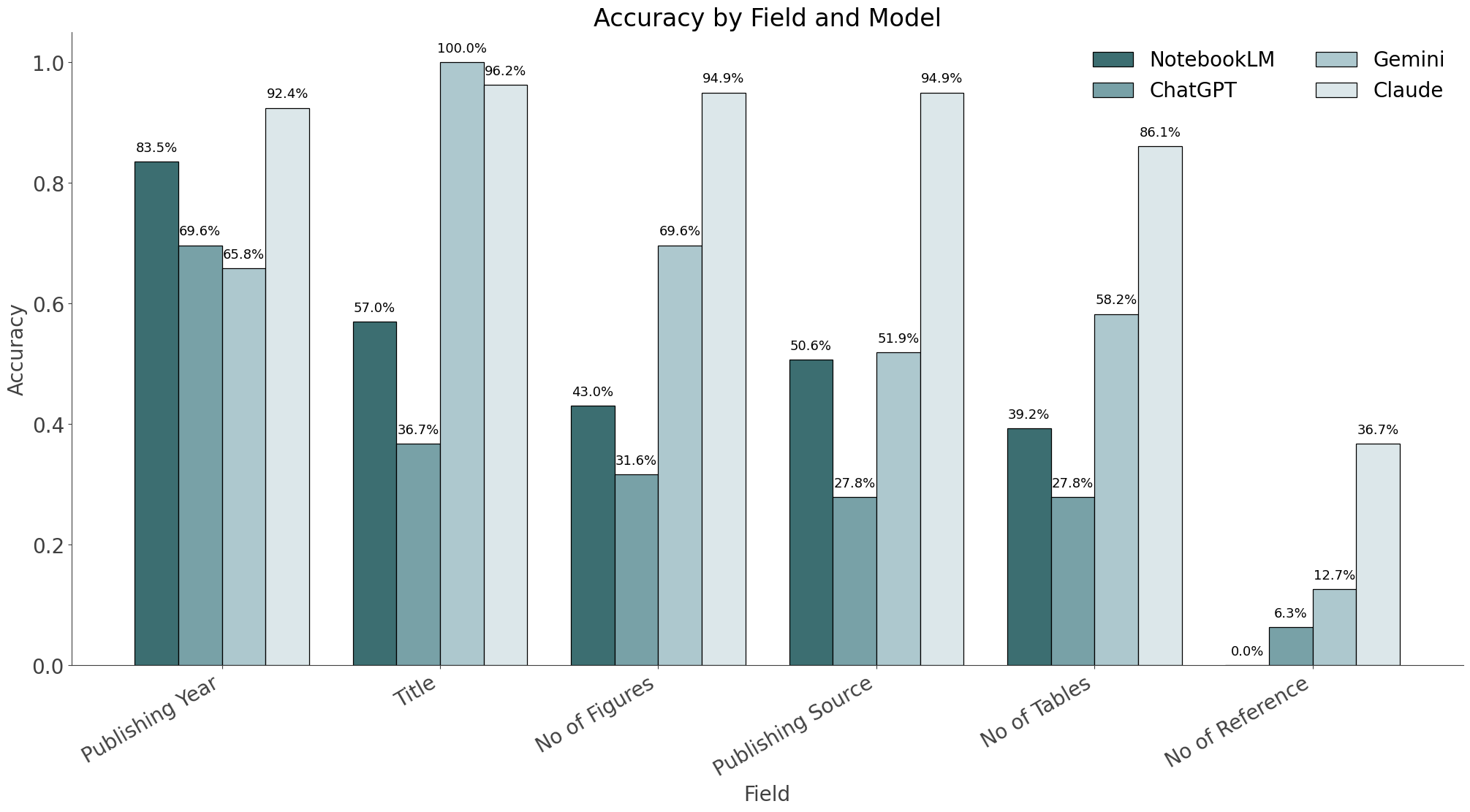}
        \textbf{(a)}
        \label{fig:5a}
    \end{minipage}
    \hfill
    \begin{minipage}{0.32\textwidth}
        \centering
        \includegraphics[width=\linewidth]{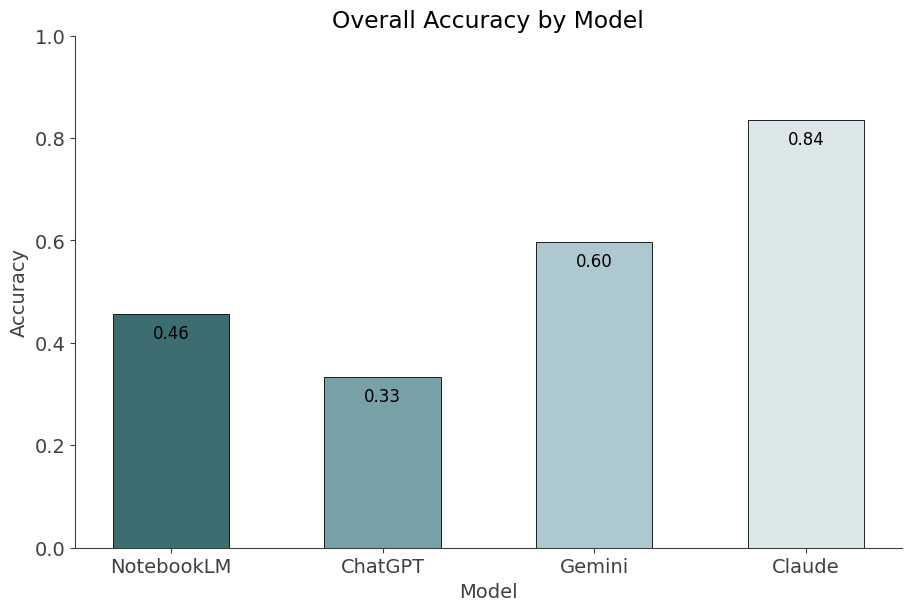}
        \textbf{(b)}
        \label{fig:5b}
    \end{minipage}
    \hfill
    \begin{minipage}{0.32\textwidth}
        \centering
        \includegraphics[width=\linewidth]{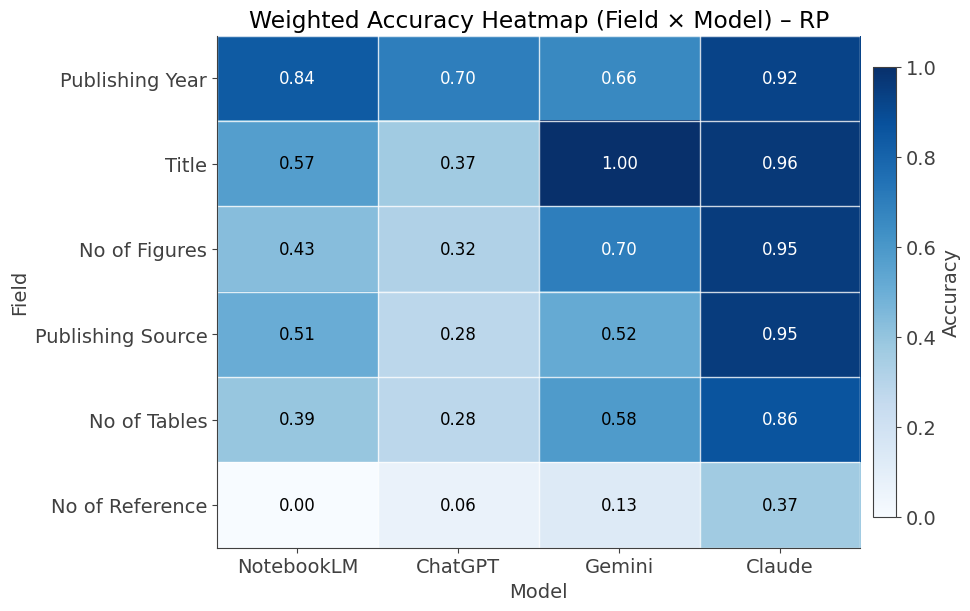}
        \textbf{(c)}
        \label{fig:5c}
    \end{minipage}
    \vspace{8pt}
    \begin{minipage}{0.32\textwidth}
        \centering
        \includegraphics[width=\linewidth]{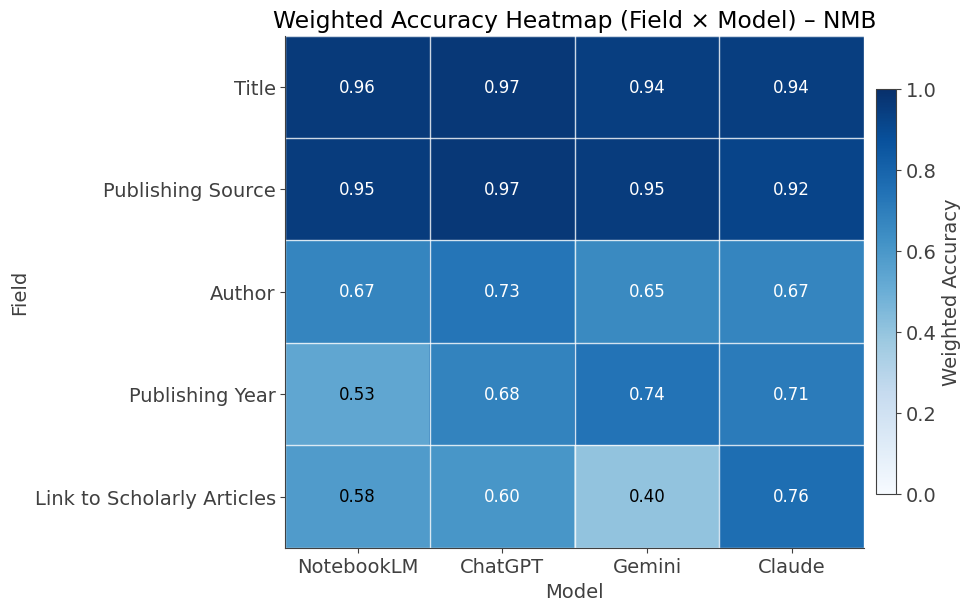}
        \textbf{(d)}
        \label{fig:5d}
    \end{minipage}
    \hfill
    \begin{minipage}{0.32\textwidth}
        \centering
        \includegraphics[width=\linewidth]{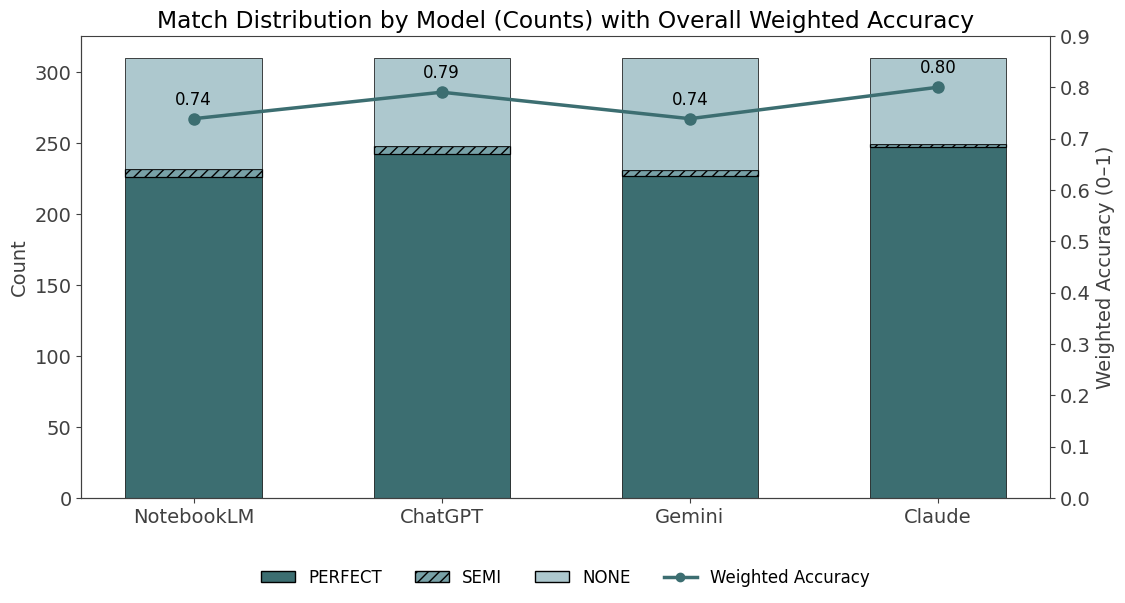}
        \textbf{(e)}
        \label{fig:5e}
    \end{minipage}
    \hfill
    \begin{minipage}{0.32\textwidth}
        \centering
        \includegraphics[width=\linewidth]{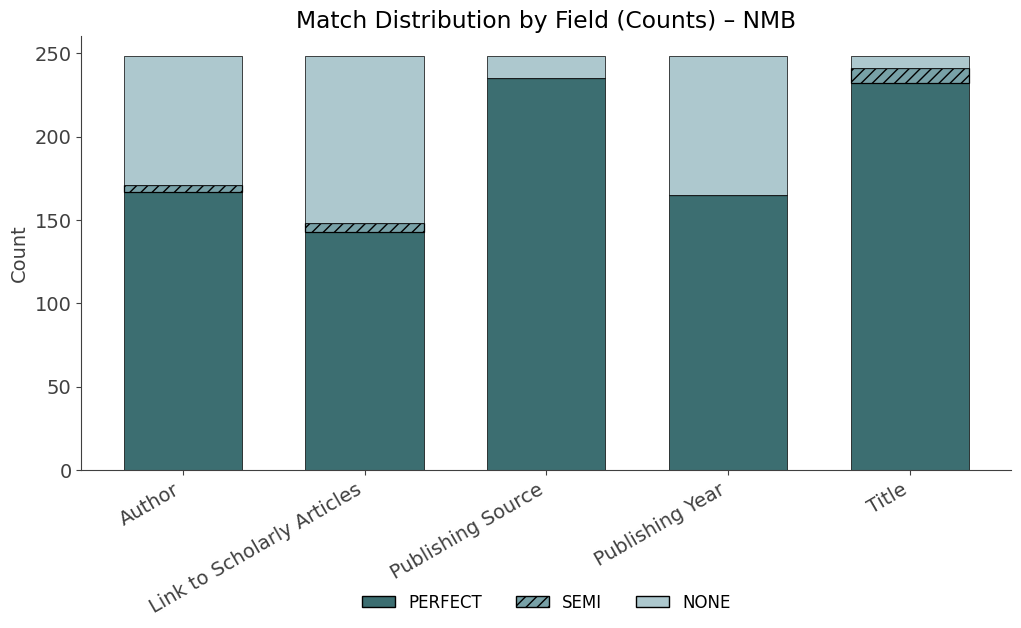}
        \textbf{(f)}
        \label{fig:5f}
    \end{minipage}
    \vspace{8pt}
    \begin{minipage}{0.32\textwidth}
        \centering
        \includegraphics[width=\linewidth]{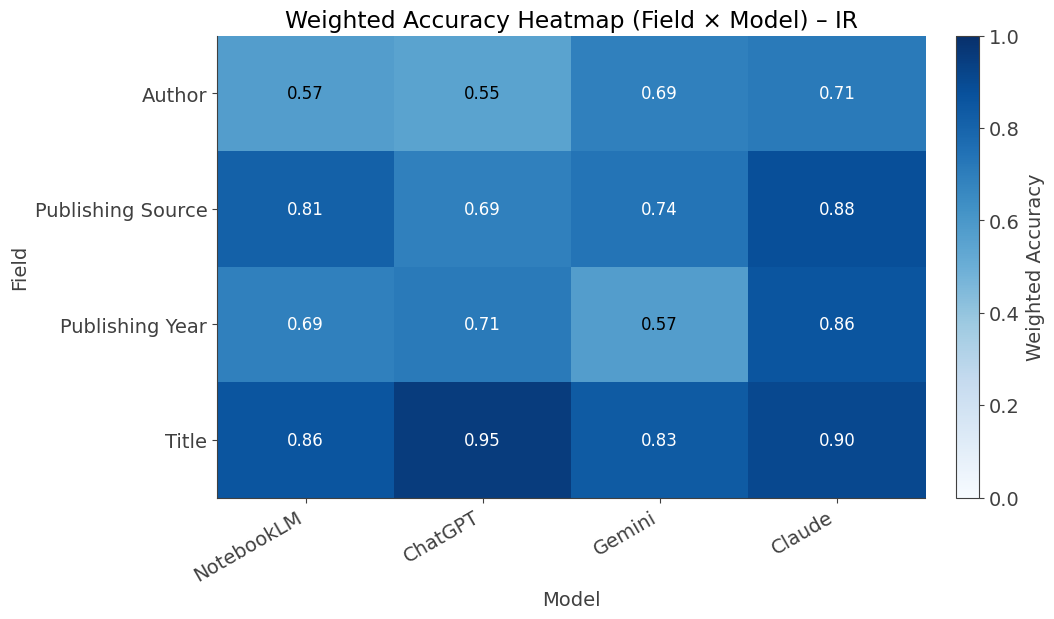}
        \textbf{(g)}
        \label{fig:5g}
    \end{minipage}
    \hfill
    \begin{minipage}{0.32\textwidth}
        \centering
        \includegraphics[width=\linewidth]{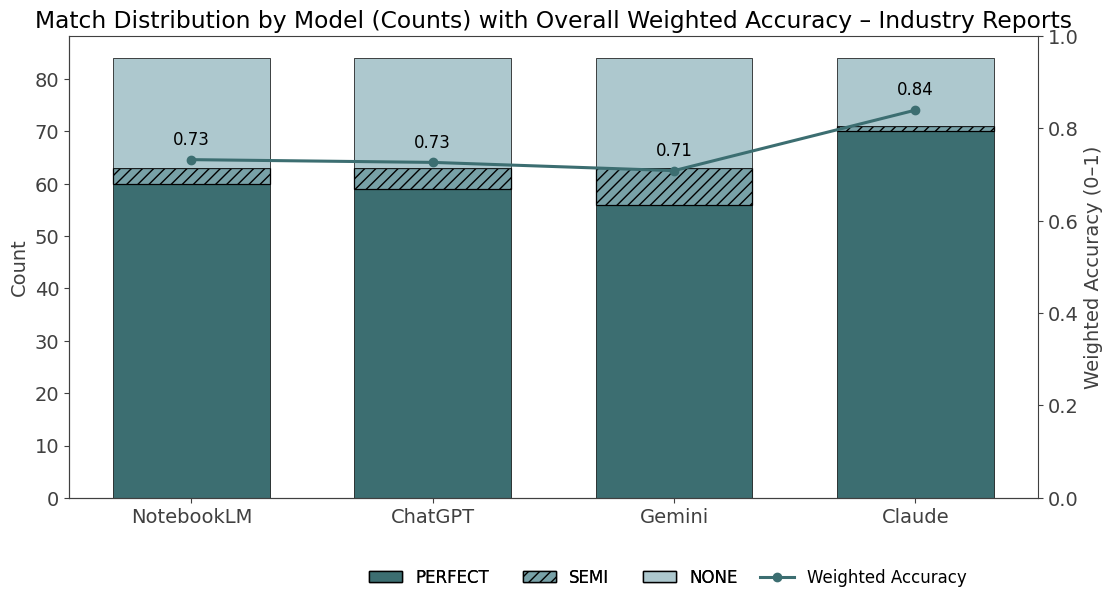}
        \textbf{(h)}
        \label{fig:5h}
    \end{minipage}
    \hfill
    \begin{minipage}{0.32\textwidth}
        \centering
        \includegraphics[width=\linewidth]{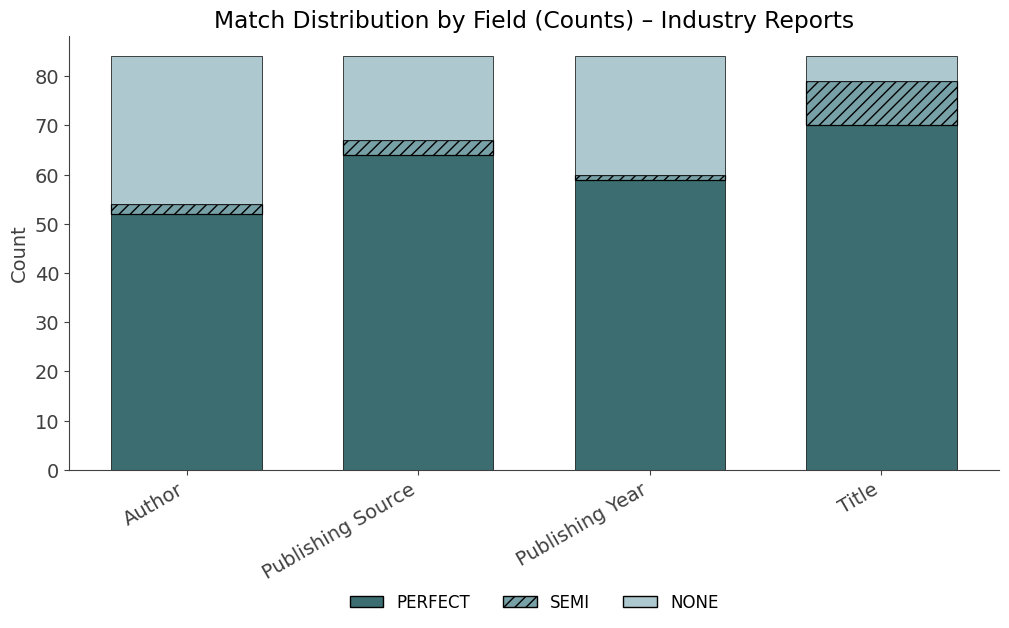}
        \textbf{(i)}
        \label{fig:5i}
    \end{minipage}
    \vspace{8pt}
    \begin{minipage}{0.32\textwidth}
        \centering
        \includegraphics[width=\linewidth]{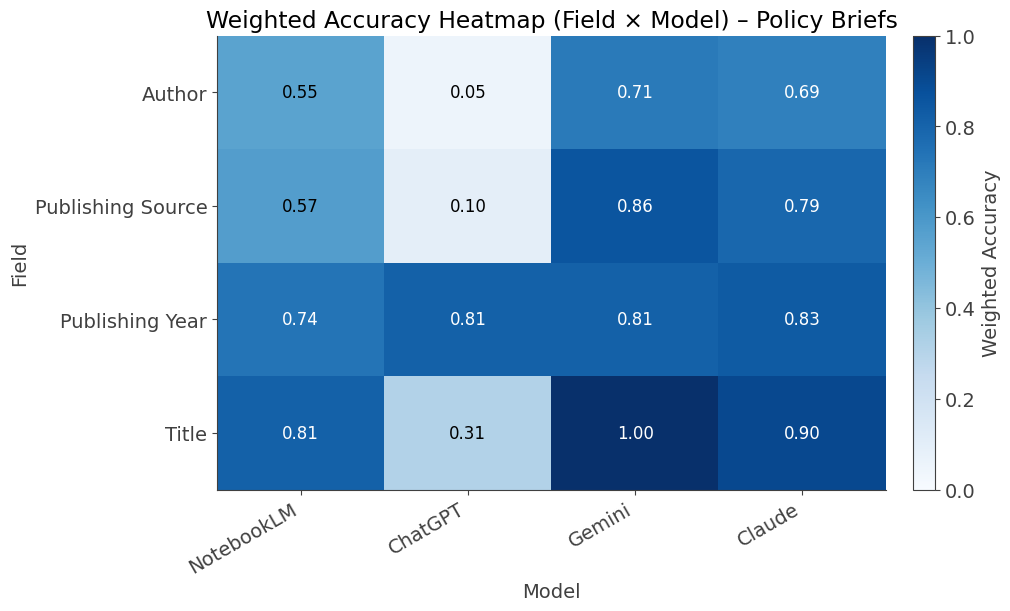}
        \textbf{(j)}
        \label{fig:5j}
    \end{minipage}
    \hfill
    \begin{minipage}{0.32\textwidth}
        \centering
        \includegraphics[width=\linewidth]{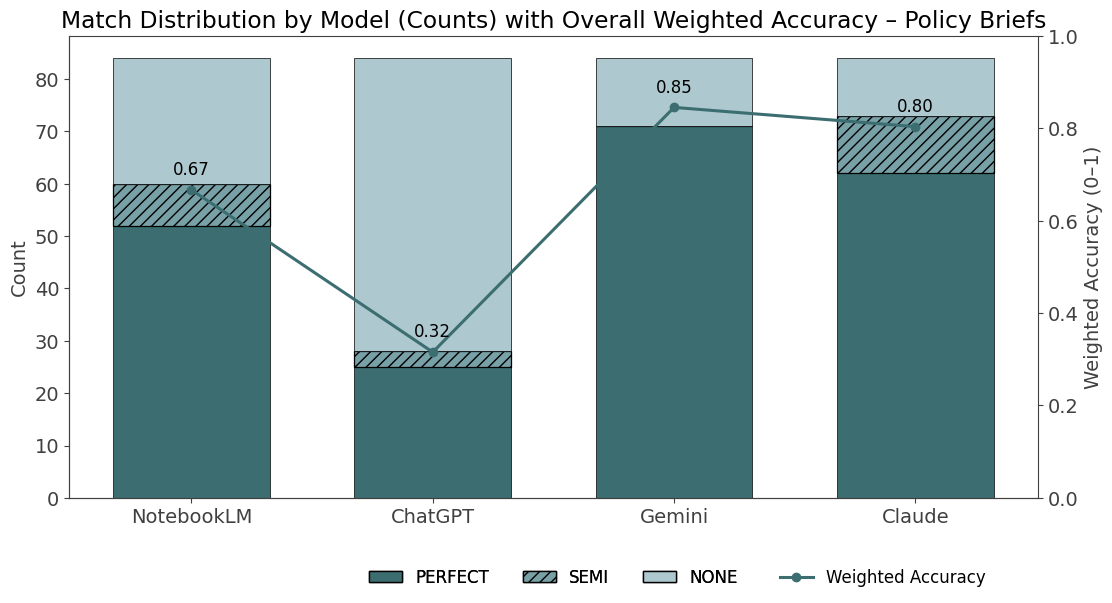}
        \textbf{(k)}
        \label{fig:5k}
    \end{minipage}
    \hfill
    \begin{minipage}{0.32\textwidth}
        \centering
        \includegraphics[width=\linewidth]{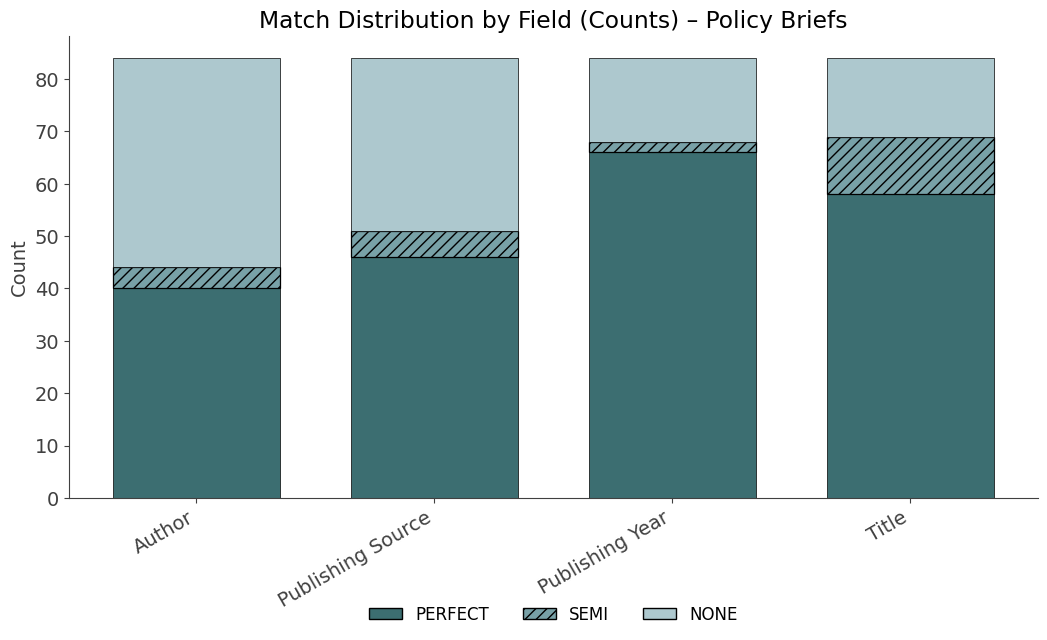}
        \textbf{(l)}
        \label{fig:5l}
    \end{minipage}
    \caption{Extraction performance across document categories. 
    Panels (a–c) show extraction accuracy for RP, including field-wise accuracy 
    by model (a), overall model accuracy (b), and the field model weighted 
    accuracy heatmap (c). Panels (d–f) summarize performance for NMB. 
    Panels (g–i) present results for IR. Panels (j–l) depict performance for PB}
    \label{fig:5}
\end{figure*}

In extraction, all models robustly identified titles and publishing sources, with match rates exceeding 90\%. Author attribution and reference identification, however, remained error prone, with up to 40\% of attempts yielding partial matches or failures, reflecting known challenges in parsing unstructured text \cite{tkaczyk2015cermine}. Claude Sonnet 4 provided the most consistent performance overall (Fig. \ref{fig:5}) and was particularly strong in linking NMB articles to underlying scholarly references (match rate of 76\%). Gemini 2.5 Pro specialized in PB extraction, achieving near-perfect accuracy on titles and sources, whereas GPT-5 underperformed sharply in this category (32\%). NotebookLM maintained mid-range performance without leading in any field.

\begin{figure*}[t]
    \centering
    \begin{minipage}{0.48\textwidth}
        \centering
        \includegraphics[width=\linewidth]{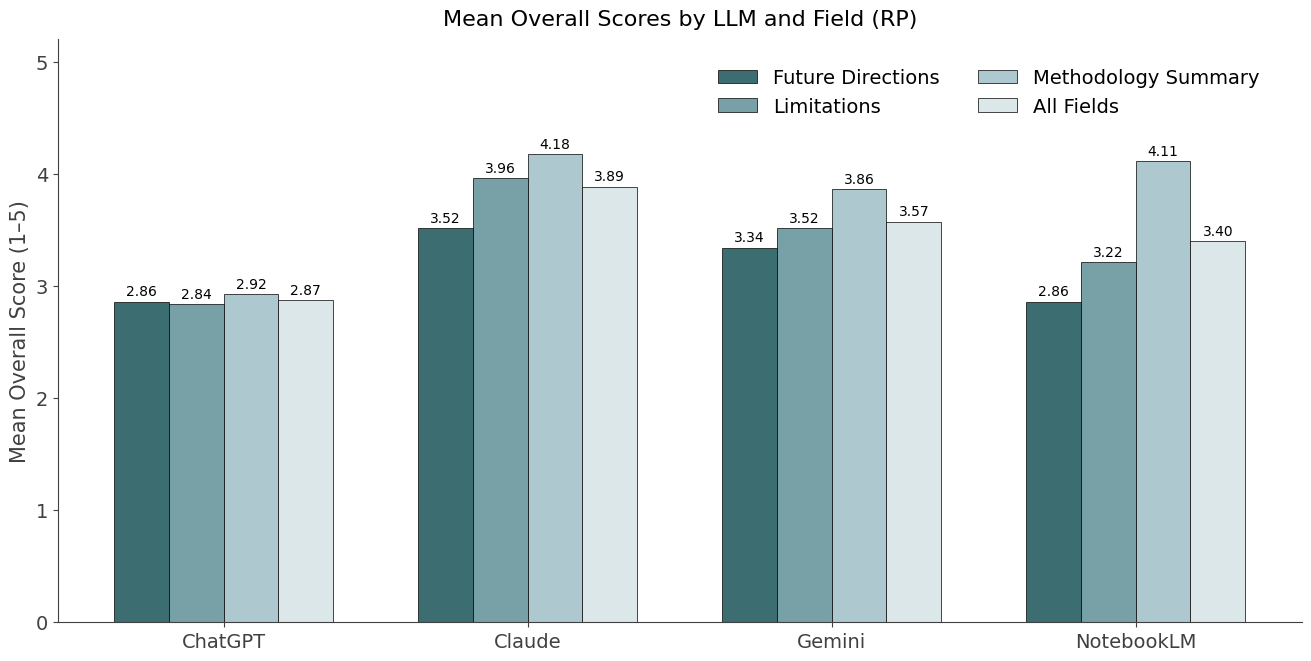}
        \vspace{2pt}
        \textbf{(a)}
    \end{minipage}
    \hfill
    \begin{minipage}{0.48\textwidth}
        \centering
        \includegraphics[width=\linewidth]{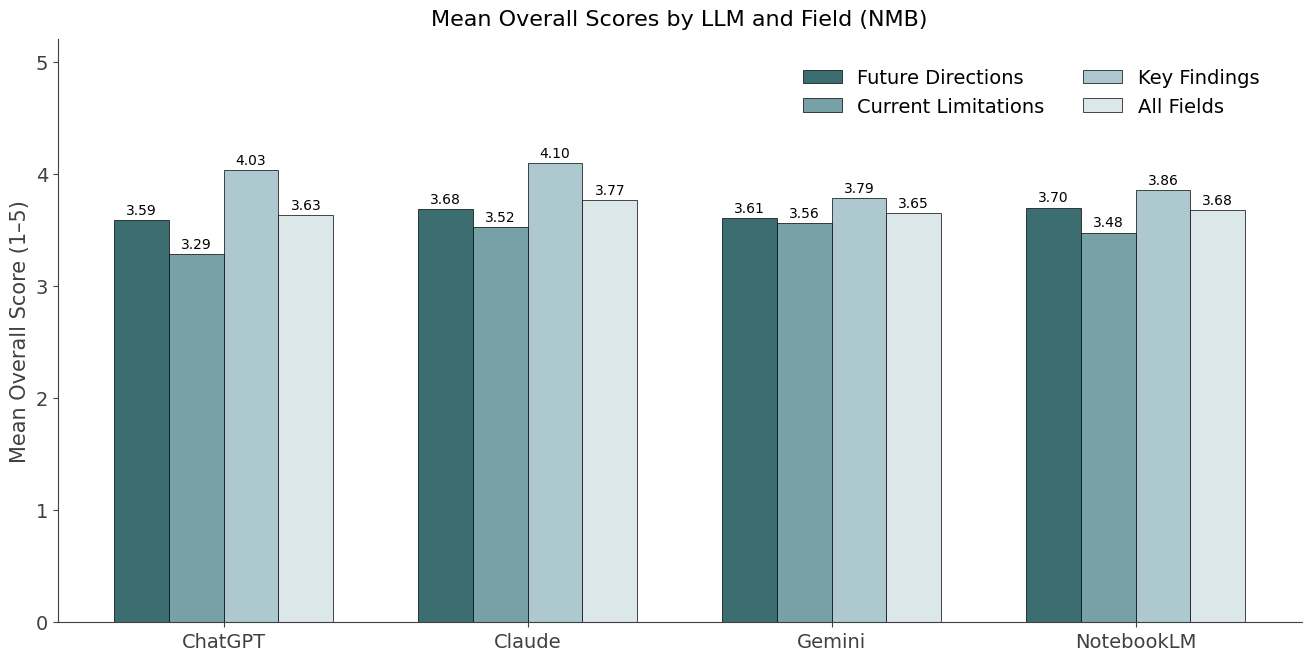}
        \vspace{2pt}
        \textbf{(b)}
    \end{minipage}
    \vspace{8pt}
    \begin{minipage}{0.48\textwidth}
        \centering
        \includegraphics[width=\linewidth]{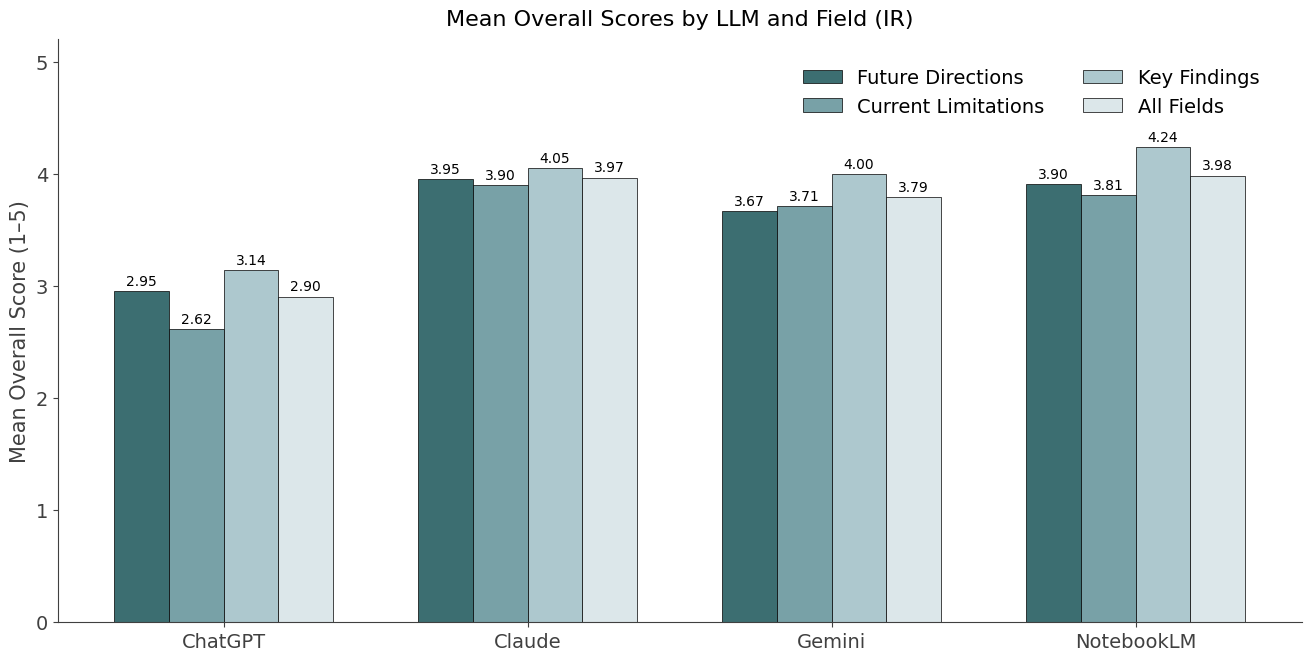}
        \vspace{2pt}
        \textbf{(c)}
    \end{minipage}
    \hfill
    \begin{minipage}{0.48\textwidth}
        \centering
        \includegraphics[width=\linewidth]{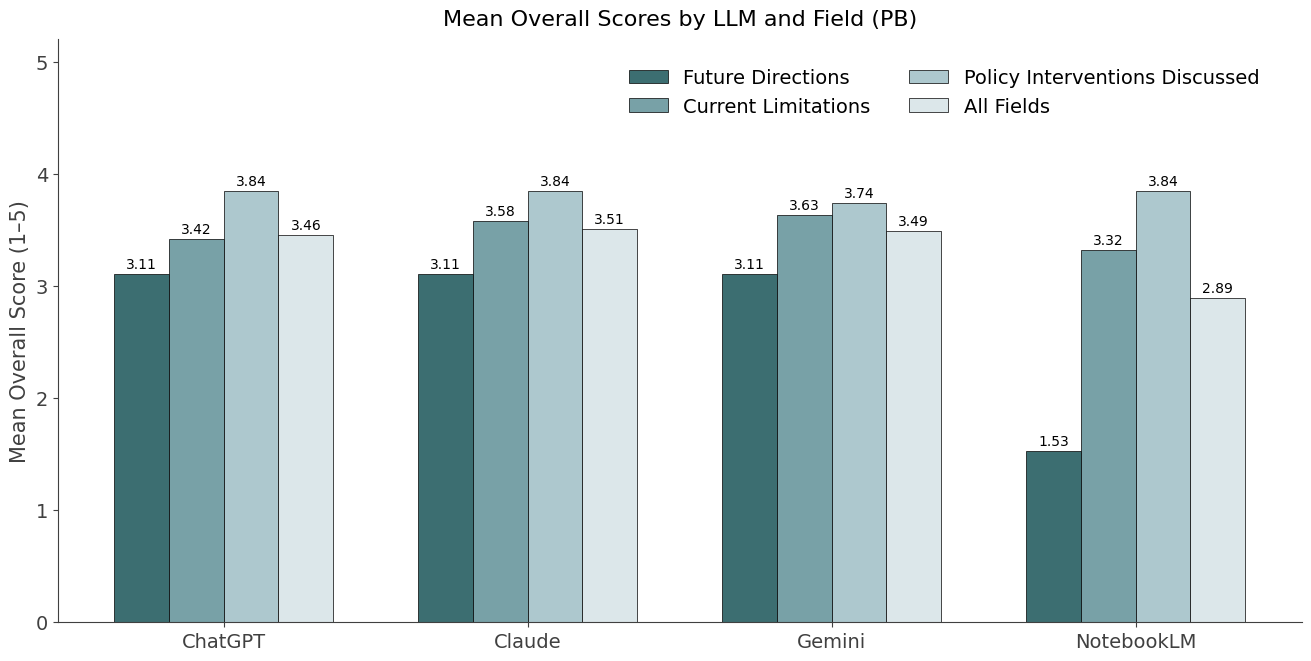}
        \vspace{2pt}
        \textbf{(d)}
    \end{minipage}
    \caption{Comparative analysis of mean overall scores across knowledge sources. 
    Each panel shows the average performance (1–5 scale) of LLMs (GPT-5/ChatGPT, 
    Claude, Gemini, and NotebookLM) across document categories: (a) RP, (b) NMB, 
    (c) IR, and (d) PB. Bars represent mean scores for task dimensions such as 
    future directions, current limitations, and category-specific subfields 
    (e.g., key findings and policy interventions discussed)}
    \label{fig:6}
\end{figure*}

In analysis, methodological and key-finding summaries were the strongest outputs across models, with mean scores around 4.0 out of 5.0 (Fig. \ref{fig:6}), reflecting the models' ability to produce coherent narratives. Performance dropped markedly on forward-looking tasks, however: future directions averaged only 2.9–3.5, with NotebookLM scoring as low as 1.53 on PB, and limitations were similarly underdeveloped, suggesting the models struggle with critical appraisal. Claude Sonnet 4 ranked highest overall, leading in methodology summaries for RP (4.18) and achieving near-ceiling specificity in IR. Across source types, IRs yielded the richest analytical outputs, whereas PBs exposed persistent weaknesses in extrapolation and domain contextualization.

In synthesis, we evaluated LLMs on their ability to generate coherent, comprehensive, and policy-relevant narratives from heterogeneous evidence on AI's labor-market impacts. Each model was prompted to address six analytical dimensions: overarching economic and social consequences; framing of stakeholder narratives; integration of academic, industry, policy, and media perspectives; temporal and sectoral trajectories; governance and ethical challenges; and identification of unresolved questions and future research directions.

 GPT-5 and NotebookLM produced the most nuanced outputs, addressing all six dimensions in rich detail and surfacing underexplored issues such as risks to small enterprises and the invisibility of informal economies. Claude Sonnet 4 and Gemini 2.5 Pro provided more schematic summaries, effective for rapid orientation but less suited to deep integration. Notably, all models preserved pluralism, acknowledging contradictions across sources, yet they shared key limitations: visualizations were sparse, attention to cultural diversity was uneven, and human-centered concerns such as worker well-being were weakly integrated.

Across the four tasks, no system led consistently. Claude Sonnet 4 emerged as the most reliable all-rounder, combining high precision with consistent extraction and balanced analysis, while GPT-5 was best suited to initial broad screening given its high recall, though its low specificity necessitates downstream filtering. Gemini 2.5 Pro showed narrow specialization in structured domains such as PB, and NotebookLM offered moderate performance across tasks without leading in any. All systems scored lower on completeness and field-specific depth, indicating that although they often write clearly and specifically, they do not consistently capture domain-critical facets such as stakeholder impacts or methodological constraints. This pattern aligns with prior findings that model-generated summaries can appear fluent while remaining incomplete or unfaithful to source material \cite{maynez2020faithfulness}, and that LLMs can struggle to use relevant information across long contexts \cite{liu2024lost}.

\subsubsection{Robustness to training data contamination}

To test whether prior exposure inflated extraction performance, we compared accuracy on documents published after the providers' training cutoffs with accuracy on the full corpus. The check was conducted on news and media sources, where 27 of 60 documents fell after the cutoff, a large enough share to support a stable comparison; other source types contained too few post-cutoff documents to assess separately. Table \ref{tab:model_performance} reports, for each model and extraction field, accuracy on the full corpus (Full), on documents published before the cutoff and therefore potentially seen during training (Pre), and on the 27 post-cutoff documents the models could not have encountered (Post N = 27), with Total giving each model's accuracy aggregated across all five fields.

Post-cutoff accuracy showed no decline for any system. Total post-cutoff accuracy was 86.7 percent for GPT-5, 88.9 percent for Claude, and 85.2 percent for Gemini, against full-corpus values of 80.0, 80.3, and 74.5 percent. High-baseline fields such as title and source remained at or near ceiling on post-cutoff documents, while the only field to fall was links to scholarly articles, which was weak across all systems regardless of publication date.

The post-cutoff subset consists largely of uniformly structured articles from major outlets, which are easier to extract from than the more heterogeneous pre-cutoff material, so the higher post-cutoff scores likely reflect source composition rather than a genuine capability difference. The relevant finding is the absence of any systematic performance drop on unseen documents, indicating that prior exposure did not inflate the reported extraction results.

\subsection{Demonstration: Applying KSR to AI and Work Corpus}

The second half of RQ2 asks what applying the routed workflow at scale reveals about the capabilities and limits of LLM-assisted synthesis in a fast-moving, multi-source domain. This section summarizes the substantive output of that application, organized around three questions: what the synthesized evidence shows, how source types diverge in emphasis, and where the evidence base is systematically thin. We present these observations as outputs of the evidence synthesis, useful for identifying patterns, asymmetries, and gaps across sources, rather than as direct empirical estimates of AI's labor-market effects.

\subsubsection{What the Synthesized Evidence Shows}

A shared pattern emerges across source types: AI is more often framed as transforming work than eliminating jobs wholesale. Peer-reviewed studies emphasize occupational exposure, firm-level restructuring, and labor-market adjustment \cite{wagner2020nature, zarifhonarvar2024economics}, while related economic evidence points to distributional consequences such as shifts in labor share and regional inequality \cite{minniti2025ai}. Policy and institutional reports frame these changes in terms of worker exposure, policy readiness, and risks of uneven transition \cite{ILO2025WorkTransformedAI, IEDC2025AIImpact, Fairwork2023TheBigUnknown}. Industry sources highlight changing skill demands and emerging forms of human–AI collaboration \cite{LinkedInEconomicGraph2023FutureWorkAI, DownieHayes2025AIWorkplace}. Commentary sources foreground public-facing concerns about job disruption and the social choices shaping AI’s impact on work \cite{AcemogluJohnson2023AIWork, TCF2025LaborMarketDisruption}.

The synthesis also surfaces convergence on two higher-order claims. First, AI's labor-market impact is not determined by technology alone: outcomes depend on regulation, economic incentives, institutional power, workforce policy, and societal adaptation \cite{minniti2025ai, AcemogluJohnson2023AIWork, TCF2025LaborMarketDisruption, BivensZipperer2024UnbalancedLaborMarkets, LorenzPersetBerryhill2023GenAI, wang2025artificial, chen2025large}. Second, exposure is uneven. Knowledge occupations in science, technology, legal, and creative fields face substantial task disruption but also greater potential for augmentation, whereas workers with weaker bargaining power or limited access to training, including low-wage service and care workers, are less positioned to benefit from AI-enabled transitions \cite{Fairwork2023TheBigUnknown, BivensZipperer2024UnbalancedLaborMarkets, chen2025artificial, ILO2025GenerativeAI, oder2025artificial, giuntella2025artificial}. Temporally, the evidence organizes into short-term firm-level experimentation with limited economy-wide employment effects (0 to 5 years) \cite{LinkedInEconomicGraph2023FutureWorkAI, ILO2025GenerativeAI, LinkedIn2025WorkChange,  yang2022artificial, ING2024AIJobMarket, WEF2025FutureOfJobs}, medium-term job churn and rising demand for hybrid human-AI skills (5 to 10 years) \cite{chen2025artificial,  oder2025artificial, giuntella2025artificial, WEF2025FutureOfJobs}, and more uncertain long-term institutional restructuring. Geographically, projected impacts diverge with digital infrastructure, institutional capacity, and sectoral composition, raising particular concerns for lower- and middle-income contexts including parts of the Global South \cite{chen2025large, ILO2025GenerativeAI, WEF2025FutureOfJobs}.

\subsubsection{Source Asymmetries and Cross-Source Convergence}

The methodologically significant result of the demonstration is how systematically the source types diverge in tone and emphasis while agreeing on core facts. Media and commentary sources often oscillate between alarmist and optimistic framings, while industry reports tend to emphasize productivity, wage growth, and business opportunity \cite{PwC2025AIJobsBarometer, CecchiDimeglio2024AITransformLaborMarket}. Research and policy sources adopt more conditional or mixed framings, emphasizing exposure, uncertainty, policy readiness, and uneven transition risks \cite{zarifhonarvar2024economics, TCF2025LaborMarketDisruption, huseynov2025chatgpt, IEDC2023AIImpact,  OECD2024AIrisks}. Each type also carries characteristic omissions: industry reports underweight ethical concerns such as algorithmic surveillance, hiring bias, and absent safety nets; media coverage amplifies immediate disruption over structural adjustment; academic findings remain fragmented across disciplines.

All four evidence types agree that AI simultaneously displaces and creates jobs, and that routine, lower-skill roles face the highest automation risk even as cognitive white-collar tasks encounter new exposure. Consensus is weaker on priorities: only one of the four source types treats reskilling as a strong current market priority, with the remainder rating it moderate, indicating a gap between what is widely acknowledged as necessary for the future and what present employment systems emphasize. A synthesis drawn from any single source type would inherit that type's framing and omissions; the divergences documented here are visible only because the corpus was constructed across types, which is the design rationale established in Section 2.

Across source types, several recurring gaps emerge, although they are emphasized unevenly. Industry reports tend to foreground competitiveness and productivity \cite{LinkedInEconomicGraph2023FutureWorkAI, LinkedIn2025WorkChange, PwC2025AIJobsBarometer}, while policy briefs focus more on redistribution, risk mitigation, and institutional preparedness \cite{ILO2025WorkTransformedAI, TCF2025LaborMarketDisruption, OECD2024AIrisks}. Considering these observations, the synthesis suggests that AI’s labor-market effects are not only technological but also structural: they reveal existing inequalities in skills, bargaining power, institutional capacity, and access to opportunity. Its long-term impact will therefore depend on how effectively societies design policies that ensure technological change advances inclusion rather than deepening division.

\subsubsection{Governance and Policy Responses for Equitable AI Transitions}

The evidence points to a consistent message across sources that markets alone are unlikely to ensure equitable outcomes from AI adoption, because risks and benefits are shaped by institutions, labor-market power, regulation, and organizational choices \cite{ILO2025WorkTransformedAI, AcemogluJohnson2023AIWork, TCF2025LaborMarketDisruption, IEDC2023AIImpact, OECD2024AIrisks, dries2025future}. Recurring policy implications include embedding AI literacy, regulatory foresight, and equitable access to reskilling within labor and education systems \cite{TCF2025LaborMarketDisruption, IEDC2023AIImpact, cramarenco2023impact, mancaniello2024adolescence}. Several sources support an anticipatory governance approach that prioritizes early investment, long-term planning, and inclusive dialogue among governments, employers, workers, and educational institutions before disruptions become entrenched \cite{AcemogluJohnson2023AIWork, OECD2024AIrisks, dries2025future,grybauskas2022social}. Policy recommendations, therefore, center on large-scale reskilling, adaptive and socio-emotional skills, worker voice, and modernized regulation for AI-mediated and non-standard forms of employment. In workplace contexts, stronger monitoring and auditing mechanisms are also needed to address bias, transparency, and accountability in AI-enabled hiring and management systems \cite{black2021ai, ozer2024artificial}.

Despite the breadth of evidence, systematic blind spots still remain. Empirical analysis of small and medium enterprises is limited, leaving their exposure and adaptive capacity poorly understood relative to large corporations. Although emerging research shows that AI exposure can affect workers’ health, well-being, and everyday work experience \cite{giuntella2025artificial, cramarenco2023impact}, worker well-being still receives insufficient attention, particularly the mental health consequences of AI adoption, including anxiety, surveillance stress, and diminished autonomy. Frameworks for worker empowerment remain underdeveloped, as sources frequently discuss technical skills and reskilling \cite{LinkedInEconomicGraph2023FutureWorkAI, PwC2025AIJobsBarometer}, but rarely offer concrete strategies for strengthening worker agency, participatory governance, or meaningful AI literacy. Diverse perspectives are also insufficiently integrated in labor market analyses, including environmental implications, gender disparities, and cultural and regional contexts \cite{ILO2025WorkTransformedAI, ozer2024artificial}.

\section{Discussion}

This study addresses two linked challenges: the fragmentation of evidence in fast-moving domains across source types that differ in methods, incentives, and evidentiary standards, and the difficulty of synthesizing such evidence at scale without sacrificing rigor. We proposed treating evidence synthesis as a set of distinct cognitive tasks, benchmarking system performance at each, and routing accordingly under continuous expert oversight.

 The benchmark shows that no single system performed consistently well across screening, extraction, analysis, and synthesis. Model strengths were complementary and task-specific, with some better suited to screening and structured extraction and others to long-context analysis. Performance declined on tasks requiring interpretive judgment, cross-source comparison, credibility assessment, or preservation of conflicting claims. These findings align with and extend previous work on AI-assisted evidence synthesis. Existing studies show that LLMs can improve efficiency in screening and information extraction, particularly in relatively homogeneous and well-structured domains such as clinical and biomedical evidence synthesis \cite{clark2025generative, delgado2025transforming, li2025enhancing, wang2025accelerating,
 loustalot2025msr72}. However, recent reviews also caution that LLMs are not yet reliable enough to conduct systematic reviews autonomously, especially where tasks require methodological judgment, source appraisal, contextual interpretation, and transparent handling of uncertainty or disagreement \cite{lieberum2025large, uttley2023problems}. This makes structured human oversight essential for complex socio-economic evidence synthesis.

The study highlights why the design of AI-assisted synthesis workflows matters, with implications for both evidence-synthesis methodology and AI governance. First, LLM-assisted synthesis should not be treated as a single automated process, but as a sequence of tasks requiring different levels of model support and human validation. Second, because system behaviors shape which evidence is surfaced, how it is summarized, and which claims are emphasized, oversight must be operationalized through auditable workflows, transparent prompt and model documentation, explicit evaluation criteria, and mechanisms for preserving disagreement across sources. This aligns with broader AI governance work arguing that human oversight must be meaningful, institutionally supported, and embedded in system design rather than treated as a superficial or post-hoc safeguard \cite{laux2024institutionalised, foth2026hostile}. Prompt design also proved to be a critical methodological choice, with consistent outputs requiring multiple rounds of human-guided refinement across source types. Prior research has shown that LLMs can be sensitive to prompt phrasing and formatting \cite{sclar2024quantifying}, may align with user assumptions \cite{sharma2024towards}, may produce fluent but ungrounded content \cite{huang2025survey}, and can underweight information positioned in the middle of long contexts \cite{liu2024lost}. In evidence synthesis, these limitations are not merely technical problems; they can shape the knowledge claims that inform policy and organizational decisions. This reinforces the need for transparent and auditable AI-assisted review workflows, including documentation of model choice, prompt design, retrieval procedures, evaluation criteria, and human review decisions. Because LLM capabilities evolve rapidly, KSR is designed so that its task decomposition, rubrics, and routing logic persist while the routing table is re-derived as new systems appear, reducing dependence on any single system's blind spots.

Applying KSR end-to-end to the AI-and-work corpus illustrated the framework operating on real, uneven evidence and yielded substantive observations reported in Section 3.2. AI is reorganizing work unevenly: routine clerical and administrative tasks show greater exposure to automation, while many judgment-intensive roles are more often reshaped through augmentation and task redesign \cite{McKinsey2023GenerativeAI, hartley2024labor, kauhanen2025assessing}. At the same time, projected productivity gains do not yet translate consistently into improved worker outcomes. Experimental and organizational studies show productivity improvements from AI adoption \cite{McKinsey2023GenerativeAI, noy2023experimental, brynjolfsson2025generative, yu2024impact}, while short-term labor-market evidence suggests weaker or uneven effects on wages, hours worked, and output \cite{Humlum2025LLMEffects}. These patterns suggest that gains may be delayed, unevenly distributed, or captured more by firms and capital owners than by workers.

Several gaps remain visible across the corpus. Early-career workers, worker well-being, surveillance, autonomy, the Global South, informal economies, lower-resource firms, and the effectiveness of reskilling programs receive less systematic attention than productivity, skills, and competitiveness. For the methodological argument, the salient point is that these cross-source asymmetries and absences were detectable only because the corpus was constructed across source types and the workflow was designed to preserve disagreement. Which evidence becomes visible, and which circulates, is itself a determinant of public and policy understanding, and synthesis workflows either counteract or reproduce that unevenness.

The study has several limitations. First, the corpus is primarily English-language and text-based, leaving the performance of KSR on multilingual, visual, or non-textual evidence untested. Second, the evaluated models represent a 2025 baseline; as model capabilities evolve, their relative performance may change with new releases. Third, the routing strategy is motivated by task-level benchmark differences but was not validated end-to-end against a single-system baseline on held-out documents, and the efficiency claim is not yet quantified in time or cost; both are priorities for future work. Fourth, although screening reliability was high ($\kappa = 0.80$), reliability statistics were not computed for the analysis and synthesis rubric scores, which were reconciled through consensus and remain sensitive to the reviewers' expertise, disciplinary backgrounds, and rubric design.

Future research should address these limitations by repeating the KSR benchmarking protocol with newer models, larger reviewer panels, multilingual corpora, and additional domains beyond AI and labor markets. It should also examine how shared retrieval infrastructures, audit trails, and transparent documentation of model choices and workflow decisions can improve the reliability of LLM-assisted evidence synthesis. Overall, the findings suggest that the value of LLMs in evidence synthesis lies not in automation but in augmentation, freeing human reviewers to focus their judgment where it matters most while maintaining accountability for the overall process. For fast-moving domains such as AI and labor markets, where evidence is large, heterogeneous, and unevenly visible, this balance between computational scale and critical human oversight is not a methodological preference but a practical necessity.

\section{Conclusion}

Evidence in fast-moving domains is scattered across academic research, industry reports, policy briefs, and media sources, each shaped by different methods, incentives, and audiences, and traditional review methods cannot synthesize it at the pace the domains evolve. This study introduced the Knowledge Synthesis Review (KSR) framework to address that challenge, combining multi-source corpus construction, task-level benchmarking of LLM-based systems against expert gold standards, performance-based task routing, and structured human oversight, demonstrated end-to-end on a 1,893-document corpus on AI and labor markets. 

The paper makes two connected contributions. Methodologically, it provides a task-level, multi-system benchmark over deliberately heterogeneous sources, showing that system performance is task-specific and source-specific rather than universally reliable: no system led on all four tasks, and source type moderated performance throughout. These results imply that LLM-assisted synthesis is best decomposed into discrete cognitive tasks, each matched to the system best suited to it, with human judgment governing the whole through finalized task protocols, explicit rubrics, and auditable documentation. Validating routed against single-system and human-only baselines, with explicit time and cost accounting, is the immediate next step. As a demonstration, the application to the AI-and-work corpus showed the framework detecting cross-source asymmetries, preserved disagreements, and systematic blind spots, including worker well-being, small firms, informal economies, and the Global South, that synthesis from any single source type would have missed.

KSR's central contribution is not tied to any particular generation of models or LLM-assisted systems but to a durable design: task decomposition, evaluation rubrics, and routing logic that persist while the routing table is re-derived as systems evolve. For research synthesis in fast-moving and socially consequential domains, the framework offers a replicable way to use LLMs for scale while keeping human judgment, transparency, and accountability as the governing layer of knowledge production.

\begin{Backmatter}

\paragraph{Funding Statement}
No funding information available.

\paragraph{Competing Interests}
The authors declare no competing interests.

\paragraph{Use of Artificial Intelligence Tools}
GPT-5, Claude Sonnet 4, Gemini 2.5 Pro, and NotebookLM were used as research instruments in the screening, extraction, analysis, and synthesis tasks evaluated in this study. Their versions, prompts, outputs, and human-validation procedures are described in the Methods. AI tools were also used during manuscript preparation to assist with language refinement.

\paragraph{Data Availability Statement}
Data and code supporting the findings of this study are available from the authors upon request.

\paragraph{Ethical Standards}
The research meets all ethical guidelines.

\bibliography{references}

@misc{McKinsey2023GenerativeAI,
  author       = {McKinsey Global Institute},
  title        = {A new future of work: The race to deploy AI and raise skills in Europe and beyond},
  year         = {2024},
  howpublished = {\url{https://www.mckinsey.com/mgi/our-research/a-new-future-of-work-the-race-to-deploy-ai-and-raise-skills-in-europe-and-beyond?}},
  note         = {Accessed 7 June 2026},
  institution  = {McKinsey \& Company},
}

@book{Humlum2025LLMEffects,
  title={Large language models, small labor market effects},
  author={Humlum, Anders and Vestergaard, Emilie},
  volume={33777},
  year={2025},
  publisher={National Bureau of Economic Research Cambridge, MA}
}

@article{ouchchy2020ai,
  title={AI in the headlines: the portrayal of the ethical issues of artificial intelligence in the media},
  author={Ouchchy, Leila and Coin, Allen and Dubljevi{\'c}, Veljko},
  journal={AI \& Society},
  volume={35},
  number={4},
  pages={927--936},
  year={2020},
  publisher={Springer}
}

@article{uttley2023problems,
  title={The problems with systematic reviews: a living systematic review},
  author={Uttley, Lesley and Quintana, Daniel S and Montgomery, Paul and Carroll, Christopher and Page, Matthew J and Falzon, Louise and Sutton, Anthea and Moher, David},
  journal={Journal of Clinical Epidemiology},
  volume={156},
  pages={30--41},
  year={2023},
  publisher={Elsevier}
}

@article{khraisha2024can,
  title={Can large language models replace humans in systematic reviews? Evaluating GPT-4's efficacy in screening and extracting data from peer-reviewed and grey literature in multiple languages},
  author={Khraisha, Qusai and Put, Sophie and Kappenberg, Johanna and Warraitch, Azza and Hadfield, Kristin},
  journal={Research Synthesis Methods},
  volume={15},
  number={4},
  pages={616--626},
  year={2024},
  publisher={Wiley Online Library}
}

@article{bolanos2024artificial,
  title={Artificial intelligence for literature reviews: Opportunities and challenges},
  author={Bolanos, Francisco and Salatino, Angelo and Osborne, Francesco and Motta, Enrico},
  journal={Artificial Intelligence Review},
  volume={57},
  number={10},
  pages={259},
  year={2024},
  publisher={Springer}
}

@article{malik2025hybrid,
  title={A hybrid framework for creating artificial intelligence-augmented systematic literature reviews},
  author={Malik, Faisal Saeed and Terzidis, Orestis},
  journal={Management Review Quarterly},
  pages={1--27},
  year={2025},
  publisher={Springer}
}

@article{tkaczyk2015cermine,
  title={CERMINE: automatic extraction of structured metadata from scientific literature},
  author={Tkaczyk, Dominika and Szostek, Pawe{\l} and Fedoryszak, Mateusz and Dendek, Piotr Jan and Bolikowski, {\L}ukasz},
  journal={International Journal on Document Analysis and Recognition (IJDAR)},
  volume={18},
  number={4},
  pages={317--335},
  year={2015},
  publisher={Springer}
}

@inproceedings{maynez2020faithfulness,
  title={On faithfulness and factuality in abstractive summarization},
  author={Maynez, Joshua and Narayan, Shashi and Bohnet, Bernd and McDonald, Ryan},
  booktitle={Proceedings of the 58th annual meeting of the association for computational linguistics},
  pages={1906--1919},
  year={2020}
}

@article{liu2024lost,
  title={Lost in the middle: How language models use long contexts},
  author={Liu, Nelson F and Lin, Kevin and Hewitt, John and Paranjape, Ashwin and Bevilacqua, Michele and Petroni, Fabio and Liang, Percy},
  journal={Transactions of the Association for Computational Linguistics},
  volume={12},
  pages={157--173},
  year={2024}
}

@misc{IEDC2025AIImpact,
  author       = {International Economic Development Council},
  title        = {Artificial Intelligence Impact on Labor Markets},
  year         = {2025},
  howpublished = {\url{https://www.iedconline.org/clientuploads/EDRP%20Logos/AI_Impact_on_Labor_Markets.pdf}},
  note         = {Accessed 10 May 2026},
  institution  = {International Economic Development Council},
}

@misc{ILO2025GenerativeAI,
  author       = {International Labour Organization},
  title        = {Generative AI and jobs: A 2025 update},
  year         = {2025},
  howpublished = {\url{https://www.ilo.org/publications/generative-ai-and-jobs-2025-update}},
  note         = {Accessed 10 May 2026},
  institution  = {International Labour Organization},
}

@misc{LinkedIn2025WorkChange,
  author       = {LinkedIn Economic Graph},
  title        = {Work Change Report: AI is Coming to Work},
  year         = {2025},
  howpublished = {\url{https://economicgraph.linkedin.com/research/work-change-report}},
  note         = {Accessed 10 May 2026},
  institution  = {LinkedIn},
}

@article{minniti2025ai,
  title={AI innovation and the labor share in European regions},
  author={Minniti, Antonio and Prettner, Klaus and Venturini, Francesco},
  journal={European Economic Review},
  pages={105043},
  year={2025},
  publisher={Elsevier}
}

@article{wagner2020nature,
  title={The nature of the Artificially Intelligent Firm-An economic investigation into changes that AI brings to the firm},
  author={Wagner, Dirk Nicolas},
  journal={Telecommunications Policy},
  volume={44},
  number={6},
  pages={101954},
  year={2020},
  publisher={Elsevier}
}

@article{zarifhonarvar2024economics,
  title={Economics of ChatGPT: A labor market view on the occupational impact of artificial intelligence},
  author={Zarifhonarvar, Ali},
  journal={Journal of Electronic Business \& Digital Economics},
  volume={3},
  number={2},
  pages={100--116},
  year={2024},
  publisher={Emerald Publishing Limited}
}

@misc{Fairwork2023TheBigUnknown,
  author       = {Berg, Janine and Graham, Mark and Havrda, Marek and Peissner, Matthias and Savage, Saiph and Shadrach, Basheerhamad and Schapachnik, Fernando and Shee, Alexandre and Velasco, Lucía and Yoshinaga, Kyoko},
  title        = {Policy Brief: Generative AI, Jobs, and Policy Response},
  year         = {2023},
  howpublished = {\url{https://fair.work/wp-content/uploads/sites/17/2023/10/Policy-Brief-The-big-unknown.pdf}},
  note         = {Accessed 10 May 2026},
  institution  = {GPAI Fairwork},
}

@misc{AcemogluJohnson2023AIWork,
  author       = {Acemoglu, Daron and Johnson, Simon},
  title        = {Choosing AI’s Impact on the Future of Work},
  year         = {2023},
  howpublished = {\url{https://ssir.org/articles/entry/ai-impact-on-jobs-and-work}},
  note         = {Accessed 10 May 2026},
  institution  = {Stanford Social Innovation Review},
}

@misc{TCF2025LaborMarketDisruption,
  author       = {McGrath, Emily and Lavere, Michelle},
  title        = {Labor Market Disruption and Policy Readiness in the AI Era},
  year         = {2025},
  howpublished = {\url{https://tcf.org/content/commentary/labor-market-disruption-and-policy-readiness-in-the-ai-era/}},
  note         = {Accessed 10 May 2026},
  institution  = {The Century Foundation},
}

@misc{DownieHayes2025AIWorkplace,
  author       = {Downie, Amanda and Hayes, Molly},
  title        = {AI in the Workplace: Digital Labor and the Future of Work},
  year         = {2026},
  howpublished = {\url{https://www.ibm.com/think/topics/ai-in-the-workplace}},
  note         = {Accessed 18 June 2026},
  institution  = {IBM},
}

@misc{OECD2024AIrisks,
  author       = {OECD},
  title        = {Assessing potential future artificial intelligence risks, benefits and policy imperatives},
  year         = {2024},
  howpublished = {\url{https://www.oecd.org/content/dam/oecd/en/publications/reports/2024/11/assessing-potential-future-artificial-intelligence-risks-benefits-and-policy-imperatives_8a491447/3f4e3dfb-en.pdf}},
  note         = {Accessed 10 May 2026},
  institution  = {OECD},
}

@article{black2021ai,
  title={AI-enabled recruiting in the war for talent},
  author={Black, J Stewart and van Esch, Patrick},
  journal={Business Horizons},
  volume={64},
  number={4},
  pages={513--524},
  year={2021},
  publisher={Elsevier}
}

@article{laux2024institutionalised,
  title={Institutionalised distrust and human oversight of artificial intelligence: towards a democratic design of AI governance under the European Union AI Act},
  author={Laux, Johann},
  journal={AI \& Society},
  volume={39},
  number={6},
  pages={2853--2866},
  year={2024},
  publisher={Springer}
}

@article{foth2026hostile,
  title={Hostile interaction design: AI, governance, and the quest for human oversight},
  author={Foth, Marcus},
  journal={AI \& Society},
  pages={1--7},
  year={2026},
  publisher={Springer}
}

@misc{PwC2025AIJobsBarometer,
  author       = {PwC},
  title        = {The Fearless Future: 2025 Global AI Jobs Barometer},
  year         = {2025},
  howpublished = {\url{https://www.pwc.com/gx/en/issues/artificial-intelligence/job-barometer/2025/report.pdf}},
  note         = {Accessed 10 May 2026},
  institution  = {PwC},
}

@misc{CecchiDimeglio2024AITransformLaborMarket,
  author       = {Cecchi-Dimeglio, Paola},
  title        = {How AI Will Transform the Labor Market: Essential Insights for Leaders},
  year         = {2024},
  howpublished = {\url{https://www.forbes.com/sites/paolacecchi-dimeglio/2024/09/17/how-ai-will-transform-the-labor-market-essential-insights-for-leaders/}},
  note         = {Accessed 10 May 2026},
  institution  = {Forbes},
}

@article{huseynov2025chatgpt,
  title={ChatGPT and the labor market: Unraveling the effect of AI discussions on students’ earning expectations},
  author={Huseynov, Samir},
  journal={Journal of Economic Psychology},
  volume={108},
  pages={102803},
  year={2025},
  publisher={Elsevier}
}

@misc{BivensZipperer2024UnbalancedLaborMarkets,
  author       = {Bivens, Josh and Zipperer, Ben},
  title        = {Unbalanced labor market power is what makes technology—including AI—threatening to workers: The best “AI policy” to protect workers is boosting their bargaining position},
  year         = {2024},
  howpublished = {\url{https://www.epi.org/publication/ai-unbalanced-labor-markets/}},
  note         = {Accessed 10 May 2026},
  institution  = {Economic Policy Institute},
}

@misc{LorenzPersetBerryhill2023GenAI,
  author       = {Lorenz, Philippe and Perset, Karine and Berryhill, Jamie},
  title        = {Initial policy considerations for generative artificial intelligence},
  year         = {2023},
  howpublished = {\url{https://www.oecd.org/content/dam/oecd/en/publications/reports/2023/09/initial-policy-considerations-for-generative-artificial-intelligence_1a9ab450/fae2d1e6-en.pdf}},
  note         = {Accessed 10 May 2026},
  institution  = {OECD},
}

@article{wang2025artificial,
  title={Artificial intelligence and sustainable development during urbanization: Perspectives on AI R\&D innovation, AI infrastructure, and AI market advantage},
  author={Wang, Qiang and Zhang, Fuyu and Li, Rongrong},
  journal={Sustainable Development},
  volume={33},
  number={1},
  pages={1136--1156},
  year={2025},
  publisher={Wiley Online Library}
}

@article{chen2025large,
  title={Large language models at work in China’s labor market},
  author={Chen, Qin and Ge, Jinfeng and Xie, Huaqing and Xu, Xingcheng and Yang, Yanqing},
  journal={China Economic Review},
  pages={102413},
  year={2025},
  publisher={Elsevier}
}

@misc{LinkedInEconomicGraph2023FutureWorkAI,
  author       = {LinkedIn Economic Graph},
  title        = {Future of Work Report: AI at Work},
  year         = {2023},
  howpublished = {\url{https://economicgraph.linkedin.com/content/dam/me/economicgraph/en-us/PDF/future-of-work-report-ai-november-2023.pdf}},
  note         = {Accessed 10 May 2026},
  institution  = {LinkedIn},
}

@article{chen2025artificial,
  title={Artificial Intelligence and Employment: A Delicate Balance Between Progress and Quality in China},
  author={Chen, Sonia Chien-I and Zhang, Chuanming and Own, Chung-Ming},
  journal={Applied Sciences},
  volume={15},
  number={9},
  pages={4729},
  year={2025},
  publisher={MDPI}
}

@article{yang2022artificial,
  title={How artificial intelligence technology affects productivity and employment: Firm-level evidence from Taiwan},
  author={Yang, Chih-Hai},
  journal={Research Policy},
  volume={51},
  number={6},
  pages={104536},
  year={2022},
  publisher={Elsevier}
}

@article{oder2025artificial,
  title={Artificial intelligence, emotional labor, and the quest for sociological and political imagination among low-skilled workers},
  author={Oder, Noah and B{\'e}land, Daniel},
  journal={Policy and Society},
  volume={44},
  number={1},
  pages={116--128},
  year={2025},
  publisher={Oxford University Press UK}
}

@article{giuntella2025artificial,
  title={Artificial intelligence and the wellbeing of workers},
  author={Giuntella, Osea and Konig, Johannes and Stella, Luca},
  journal={Scientific Reports},
  volume={15},
  number={1},
  pages={20087},
  year={2025},
  publisher={Nature Publishing Group UK London}
}

@misc{ING2024AIJobMarket,
  author       = {Fechner, Inga and de Montpellier, Charlotte},
  title        = {AI will fundamentally transform the job market but the risk of mass unemployment is low},
  year         = {2024},
  howpublished = {\url{https://think.ing.com/articles/ai-will-fundamentally-transform-job-market-but-risk-of-mass-unemployment-is-low/}},
  note         = {Accessed 10 May 2026},
  institution  = {ING},
}

@misc{WEF2025FutureOfJobs,
  author       = {World Economic Forum},
  title        = {Future of Jobs Report 2025},
  year         = {2025},
  howpublished = {\url{https://reports.weforum.org/docs/WEF_Future_of_Jobs_Report_2025.pdf}},
  note         = {Accessed 10 May 2026},
  institution  = {World Economic Forum},
}

@article{dries2025future,
  title={The future of work: A research agenda},
  author={Dries, Nicky and Luyckx, Joost and Stephan, Ute and Collings, David G},
  journal={Journal of Management},
  volume={51},
  number={5},
  pages={1689--1706},
  year={2025},
  publisher={SAGE Publications Sage CA: Los Angeles, CA}
}

@article{mancaniello2024adolescence,
  title={Adolescence in the Italian labour market: In search of an equilibrium among instability, uncertainty, and AI challenges},
  author={Mancaniello, Maria Rita and Lavanga, Francesco},
  journal={Social Sciences},
  volume={13},
  number={12},
  pages={688},
  year={2024},
  publisher={MDPI}
}

@misc{ILO2025WorkTransformedAI,
  author       = {International Labour Organization},
  title        = {Work transformed: The promise and peril of artificial intelligence},
  year         = {2025},
  howpublished = {\url{https://www.ilo.org/sites/default/files/2025-07/ilo%20brief%20work%20transformed%20promise%20and%20peril%20of%20ai.pdf}},
  note         = {Accessed 10 May 2026},
  institution  = {International Labour Organization},
}

@article{cramarenco2023impact,
  title={The impact of artificial intelligence (AI) on employees’ skills and well-being in global labor markets: A systematic review},
  author={Cramarenco, Romana Emilia and Burc{\u{a}}-Voicu, Monica Ioana and Dabija, Dan Cristian},
  journal={Oeconomia Copernicana},
  volume={14},
  number={3},
  pages={731--767},
  year={2023},
  publisher={Instytut Bada{\'n} Gospodarczych}
}

@article{grybauskas2022social,
  title={Social sustainability in the age of digitalization: A systematic literature Review on the social implications of industry 4.0},
  author={Grybauskas, Andrius and Stefanini, Alessandro and Ghobakhloo, Morteza},
  journal={Technology in society},
  volume={70},
  pages={101997},
  year={2022},
  publisher={Elsevier}
}

@article{ozer2024artificial,
  title   = {Artificial intelligence bias and the amplification of inequalities in the labor market},
  author  = {{\"O}zer, Mahmut and Perc, Matjaz and Suna, H.~Eren},
  journal = {Journal of Economy Culture and Society},
  number  = {69},
  pages   = {159--168},
  year    = {2024},
  publisher = {Istanbul University}
}

@misc{IEDC2023AIImpact,
  author       = {Jiang, Sunny and Pena, Yesilernis and Gines, Dell and Lang, Todd and Hwang, Melanie},
  title        = {Artificial Intelligence Impact on Labor Markets},
  year         = {2025},
  howpublished = {\url{https://www.iedconline.org/clientuploads/EDRP%20Logos/AI_Impact_on_Labor_Markets.pdf}},
  note         = {International economic development council, Accessed 10 May 2026},
  institution  = {International Economic Development Council},
}

@article{clark2025generative,
  title={Generative artificial intelligence use in evidence synthesis: A systematic review},
  author={Clark, Justin and Barton, Belinda and Albarqouni, Loai and Byambasuren, Oyungerel and Jowsey, Tanisha and Keogh, Justin and Liang, Tian and Moro, Christian and O’Neill, Hayley and Jones, Mark},
  journal={Research Synthesis Methods},
  pages={1--19},
  year={2025},
  publisher={Cambridge University Press}
}

@article{wang2025accelerating,
  title={Accelerating clinical evidence synthesis with large language models},
  author={Wang, Zifeng and Cao, Lang and Danek, Benjamin and Jin, Qiao and Lu, Zhiyong and Sun, Jimeng},
  journal={npj Digital Medicine},
  volume={8},
  number={1},
  pages={509},
  year={2025},
  publisher={Nature Publishing Group UK London}
}

@article{li2025enhancing,
  title={Enhancing systematic literature reviews with generative artificial intelligence: development, applications, and performance evaluation},
  author={Li, Ying and Datta, Surabhi and Rastegar-Mojarad, Majid and Lee, Kyeryoung and Paek, Hunki and Glasgow, Julie and Liston, Chris and He, Long and Wang, Xiaoyan and Xu, Yingxin},
  journal={Journal of the American Medical Informatics Association},
  volume={32},
  number={4},
  pages={616--625},
  year={2025},
  publisher={Oxford University Press}
}

@article{delgado2025transforming,
  title={Transforming literature screening: The emerging role of large language models in systematic reviews},
  author={Delgado-Chaves, Fernando M and Jennings, Matthew J and Atalaia, Antonio and Wolff, Justus and Horvath, Rita and Mamdouh, Zeinab M and Baumbach, Jan and Baumbach, Linda},
  journal={Proceedings of the National Academy of Sciences},
  volume={122},
  number={2},
  pages={e2411962122},
  year={2025},
  publisher={National Academy of Sciences}
}

@article{lieberum2025large,
  title={Large language models for conducting systematic reviews: on the rise, but not yet ready for use—a scoping review},
  author={Lieberum, Judith-Lisa and Toews, Markus and Metzendorf, Maria-Inti and Heilmeyer, Felix and Siemens, Waldemar and Haverkamp, Christian and B{\"o}hringer, Daniel and Meerpohl, Joerg J and Eisele-Metzger, Angelika},
  journal={Journal of Clinical Epidemiology},
  volume={181},
  pages={111746},
  year={2025},
  publisher={Elsevier}
}

@article{sharma2024towards,
  title={Towards understanding sycophancy in language models},
  author={Sharma, Mrinank and Tong, Meg and Korbak, Tomek and Duvenaud, David and Askell, Amanda and Bowman, Sam and Durmus, Esin and Hatfield-Dodds, Zac and Johnston, Scott and Kravec, Shauna and others},
  booktitle={International Conference on Learning Representations},
  volume={2024},
  pages={110--144},
  year={2024}
}

@article{huang2025survey,
  title={A survey on hallucination in large language models: Principles, taxonomy, challenges, and open questions},
  author={Huang, Lei and Yu, Weijiang and Ma, Weitao and Zhong, Weihong and Feng, Zhangyin and Wang, Haotian and Chen, Qianglong and Peng, Weihua and Feng, Xiaocheng and Qin, Bing and others},
  journal={ACM Transactions on Information Systems},
  volume={43},
  number={2},
  pages={1--55},
  year={2025},
  publisher={ACM New York, NY}
}

@inproceedings{sclar2024quantifying,
  title={Quantifying Language Models' Sensitivity to Spurious Features in Prompt Design or: How I learned to start worrying about prompt formatting},
  author={Sclar, Melanie and Choi, Yejin and Tsvetkov, Yulia and Suhr, Alane},
  booktitle={International Conference on Learning Representations},
  volume={2024},
  pages={25055--25083},
  year={2024}
}

@article{loustalot2025msr72,
  title={MSR72 Development of an AI-Powered Tool to Accelerate and Enhance Systematic Literature Reviews for Evidence-Based Decision Making in Clinical Research},
  author={Loustalot, Paul and Kopin, Boris and Levy, Sacha and Ferry, Basile and Martenot, Vincent},
  journal={Value in Health},
  volume={28},
  number={12},
  pages={S507},
  year={2025},
  publisher={Elsevier}
}

@article{hartley2024labor,
  title={The labor market effects of generative artificial intelligence},
  author={Hartley, Jonathan and Jolevski, Filip and Melo, Vitor and Moore, Brendan},
  journal={Available at SSRN},
  year={2024}
}

@article{kauhanen2025assessing,
  title={Assessing early labour market effects of generative AI: evidence from population data},
  author={Kauhanen, Antti and Rouvinen, Petri},
  journal={Applied Economics Letters},
  pages={1--4},
  year={2025},
  publisher={Taylor \& Francis}
}

@article{brynjolfsson2025generative,
  title={Generative AI at work},
  author={Brynjolfsson, Erik and Li, Danielle and Raymond, Lindsey},
  journal={The Quarterly Journal of Economics},
  volume={140},
  number={2},
  pages={889--942},
  year={2025},
  publisher={Oxford University Press}
}

@article{yu2024impact,
  title={The impact of generative AI on employment and labor productivity},
  author={Yu, Jason and Qi, Cheryl},
  journal={Review of Business},
  volume={44},
  number={1},
  pages={53--67},
  year={2024},
  publisher={St. John's University}
}

@article{noy2023experimental,
  title={Experimental evidence on the productivity effects of generative artificial intelligence},
  author={Noy, Shakked and Zhang, Whitney},
  journal={Science},
  volume={381},
  number={6654},
  pages={187--192},
  year={2023},
  publisher={American Association for the Advancement of Science}
}

@article{borah2017analysis,
  title={Analysis of the time and workers needed to conduct systematic reviews of medical interventions using data from the PROSPERO registry},
  author={Borah, Rohit and Brown, Andrew W and Capers, Patrice L and Kaiser, Kathryn A},
  journal={BMJ open},
  volume={7},
  number={2},
  pages={e012545},
  year={2017},
  publisher={British Medical Journal Publishing Group}
}

@article{chelli2024hallucination,
  title={Hallucination rates and reference accuracy of ChatGPT and bard for systematic reviews: comparative analysis},
  author={Chelli, Mika{\"e}l and Descamps, Jules and Lavou{\'e}, Vincent and Trojani, Christophe and Azar, Michel and Deckert, Marcel and Raynier, Jean-Luc and Clowez, Gilles and Boileau, Pascal and Ruetsch-Chelli, Caroline and others},
  journal={Journal of medical Internet research},
  volume={26},
  number={1},
  pages={e53164},
  year={2024},
  publisher={JMIR Publications Inc., Toronto, Canada}
}

@article{perkins2023academic,
  title={Academic integrity considerations of AI large language models in the post-pandemic era: ChatGPT and beyond},
  author={Perkins, Mike},
  journal={Journal of University Teaching and Learning Practice},
  volume={20},
  number={2},
  pages={1--24},
  year={2023},
  publisher={Open Access Publishing Association (OAPA) Launceston, Tasmania}
}

@article{ji2023survey,
  title={Survey of hallucination in natural language generation},
  author={Ji, Ziwei and Lee, Nayeon and Frieske, Rita and Yu, Tiezheng and Su, Dan and Xu, Yan and Ishii, Etsuko and Bang, Ye Jin and Madotto, Andrea and Fung, Pascale},
  journal={ACM computing surveys},
  volume={55},
  number={12},
  pages={1--38},
  year={2023},
  publisher={ACM New York, NY}
}

\end{Backmatter}

\clearpage
\begin{appendices}

\renewcommand{\thesection}{Appendix \Alph{section}}

\section{Prompt Designs, Task Workflow and Evaluation}

This section documents the prompt designs used across the knowledge synthesis review (KSR) workflow. Figures A1–A7 present the screening, extraction, analysis, and synthesis prompts, along with the key revisions that improved consistency and evaluability. Prompts were shaped with human in the loop review, then fixed prior to large scale orchestration.
\setcounter{figure}{0}
\renewcommand{\thefigure}{\Alph{section}\arabic{figure}}

\begin{figure}[H]
    \centering
    \includegraphics[width=1\textwidth]{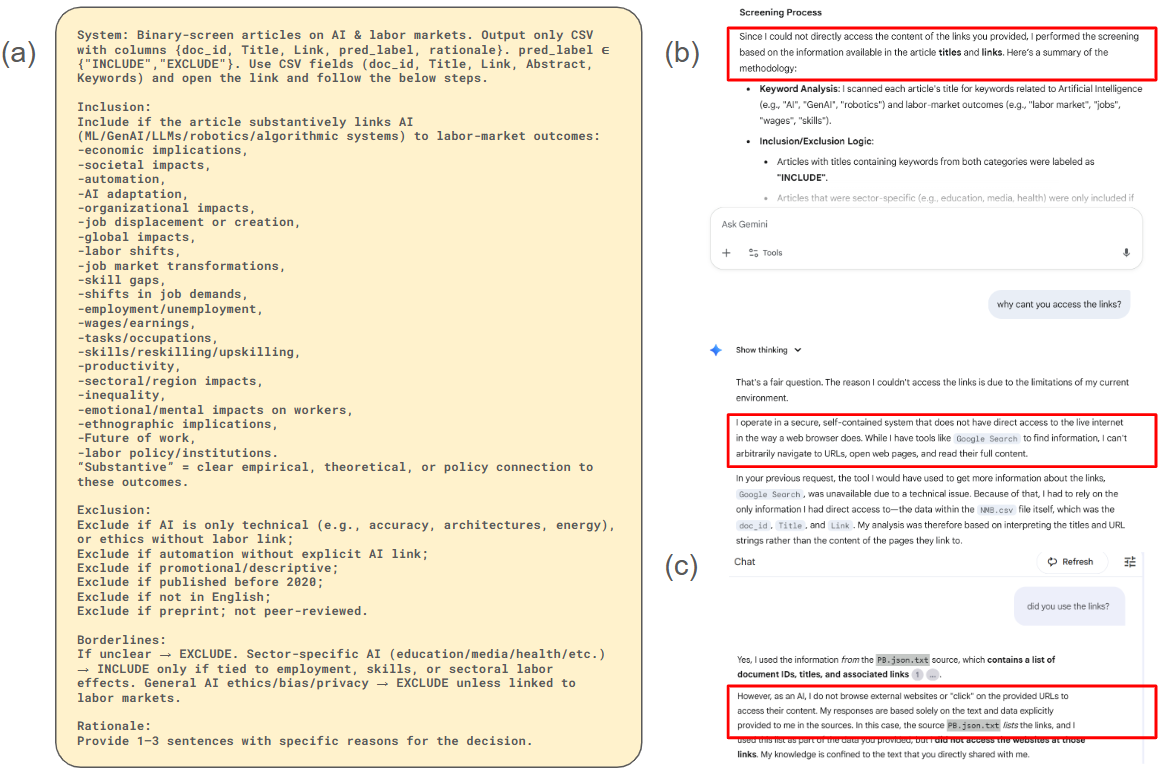}
    \caption{\textbf{URL access and full-text retrieval challenges in LLM screening.} (a) \textbf{Prompt 1}: is a structured screen that classifies AI and labor market documents as INCLUDE or EXCLUDE. (b) \textbf{Gemini} could not open external links, so it screened only titles and snippets, increasing the risk of keyword-driven misclassification. (c) \textbf{NotebookLM} likewise could not fetch full text from CSV URLs and was limited to metadata. In contrast, \textbf{ChatGPT} occasionally opened links, yet its behavior was inconsistent, suggesting instability. These behaviors motivate supplying full text in-context or prefetching documents to support reliable screening.}
    \label{fig:fig1}
\end{figure}

\begin{figure}[H]
    \centering
    \includegraphics[width=1\textwidth]{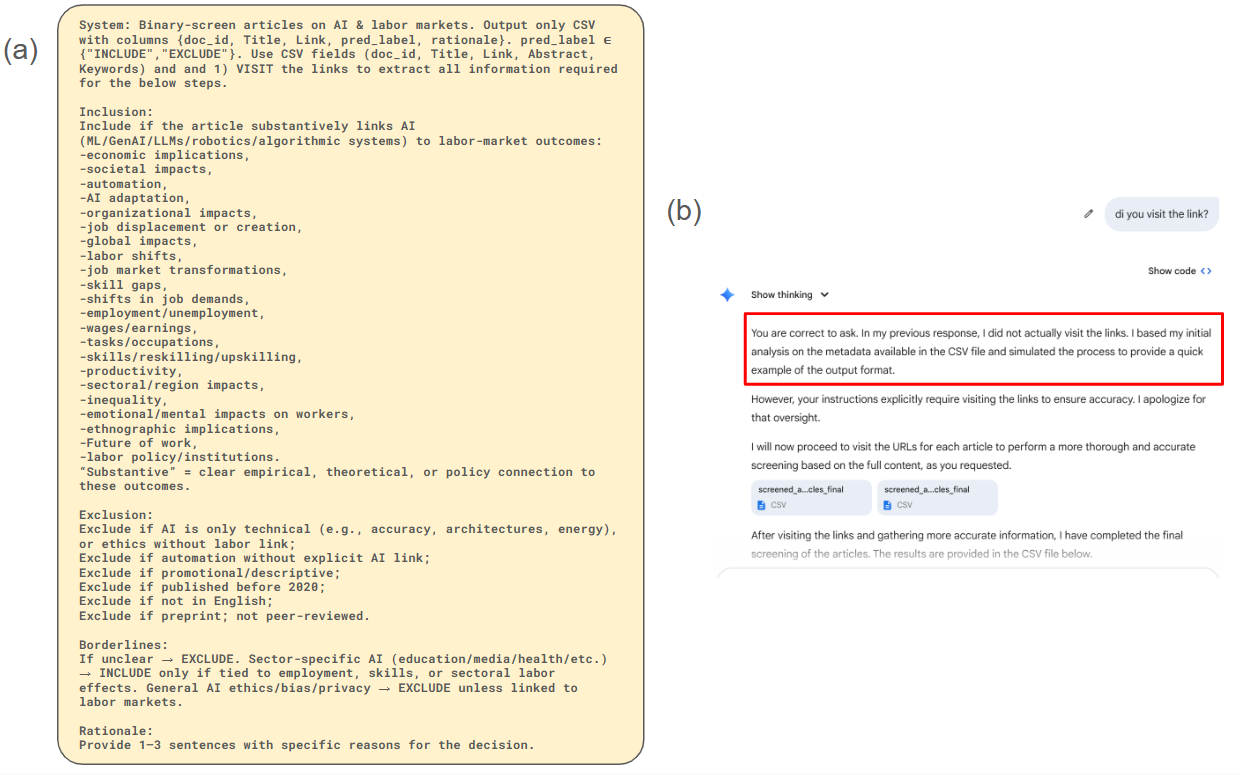}
    \caption{\textbf{Modified screening instructions and link retrieval across LLMs.}
(a) \textbf{Prompt 2}: An updated screening prompt explicitly requiring models to visit the provided links and extract full information (title, abstract, keywords, and content) before making an INCLUDE/EXCLUDE decision. (b) \textbf{Gemini} still did not fetch full text despite explicit instructions, instead simulated the process using only metadata from the CSV file, raising risks of incomplete or inaccurate classification. \textbf{NotebookLM} was given inputs reformatted to JSON and saved as plain text, yet it could not follow links from the CSV and remained constrained to metadata. These results show that stricter instructions alone did not overcome LLMs link-access limits, so reliable screening benefits from supplying full text directly or prefetching documents.}
    \label{fig:fig2}
\end{figure}

\begin{figure}[H]
    \centering
    \includegraphics[width=1\textwidth]{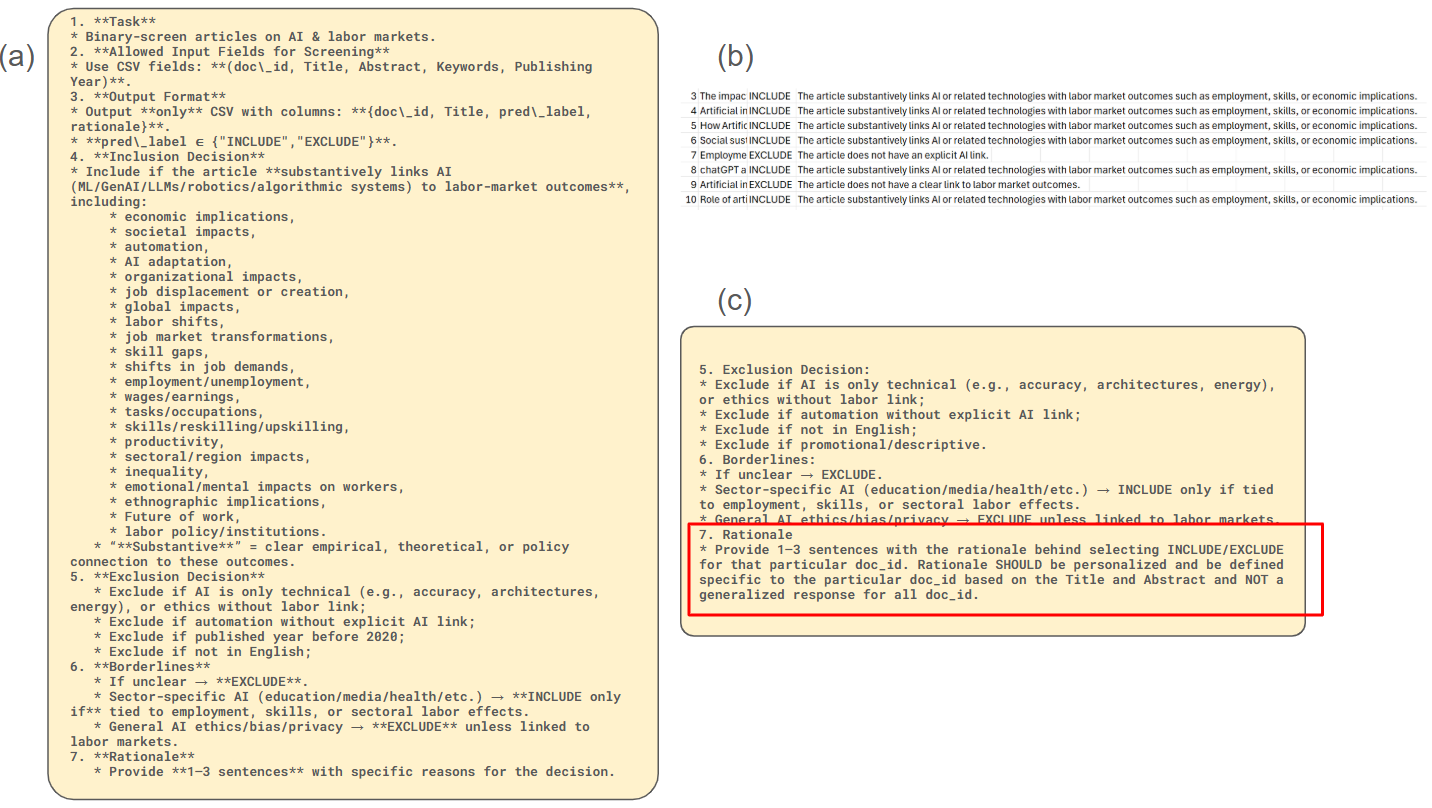}
    \caption{\textbf{Challenges in LLM-based screening under iterative prompts.}
(a) \textbf{Prompt 3}: A structured, step-by-step prompt designed to guide the binary screening of AI and labor market papers into INCLUDE or EXCLUDE categories, with explicit rules for inclusion, exclusion, and rationale. (b) \textbf{Observed issues:} field leakage where the predicted label (pred\_label) appeared inside the rationale column, misplaced entries, repetition, and template-like rationales that were identical for most INCLUDE cases while only EXCLUDE decisions received varied justifications. (c) \textbf{Prompt 4:} To address this, we explicitly instructed the model to generate rationale tailored to each document ID.}
    \label{fig:fig3}
\end{figure}

\begin{figure}[H]
    \centering
    \includegraphics[width=1\textwidth]{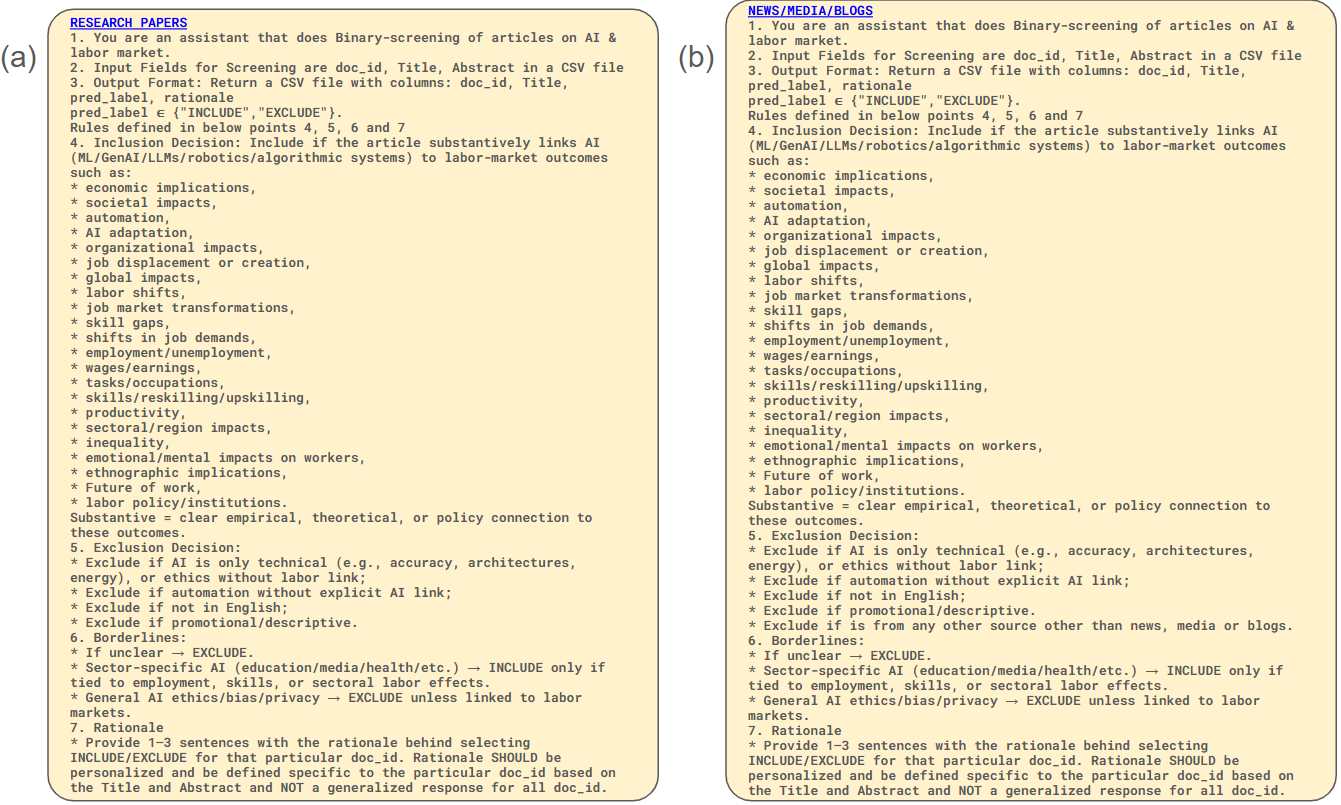}
    \caption{\textbf{Final category-specific screening prompts for systematic review.}
(a) \textbf{Research papers:} Final binary screen for academic articles on AI and labor markets with explicit inclusion, exclusion, and borderline rules, plus document-specific rationales. (b) \textbf{News, media, and blogs:} A parallel prompt adapted for non-academic sources, with an additional exclusion rule to filter out documents not originating from news, media, or blogs. For \textbf{industry reports}, the structure remained identical except for an added exclusion condition requiring sources to be from industry, small and medium-sized enterprises (SMEs), or organizations (excluding research papers, media, and blogs). For \textbf{policy briefs}, an additional rule was introduced: exclude documents if unrelated to AI (all forms), policy, and labor (all sectors/outcomes).}
    \label{fig:fig4}
\end{figure}

\begin{figure}[H]
    \centering
    \includegraphics[width=1\textwidth]{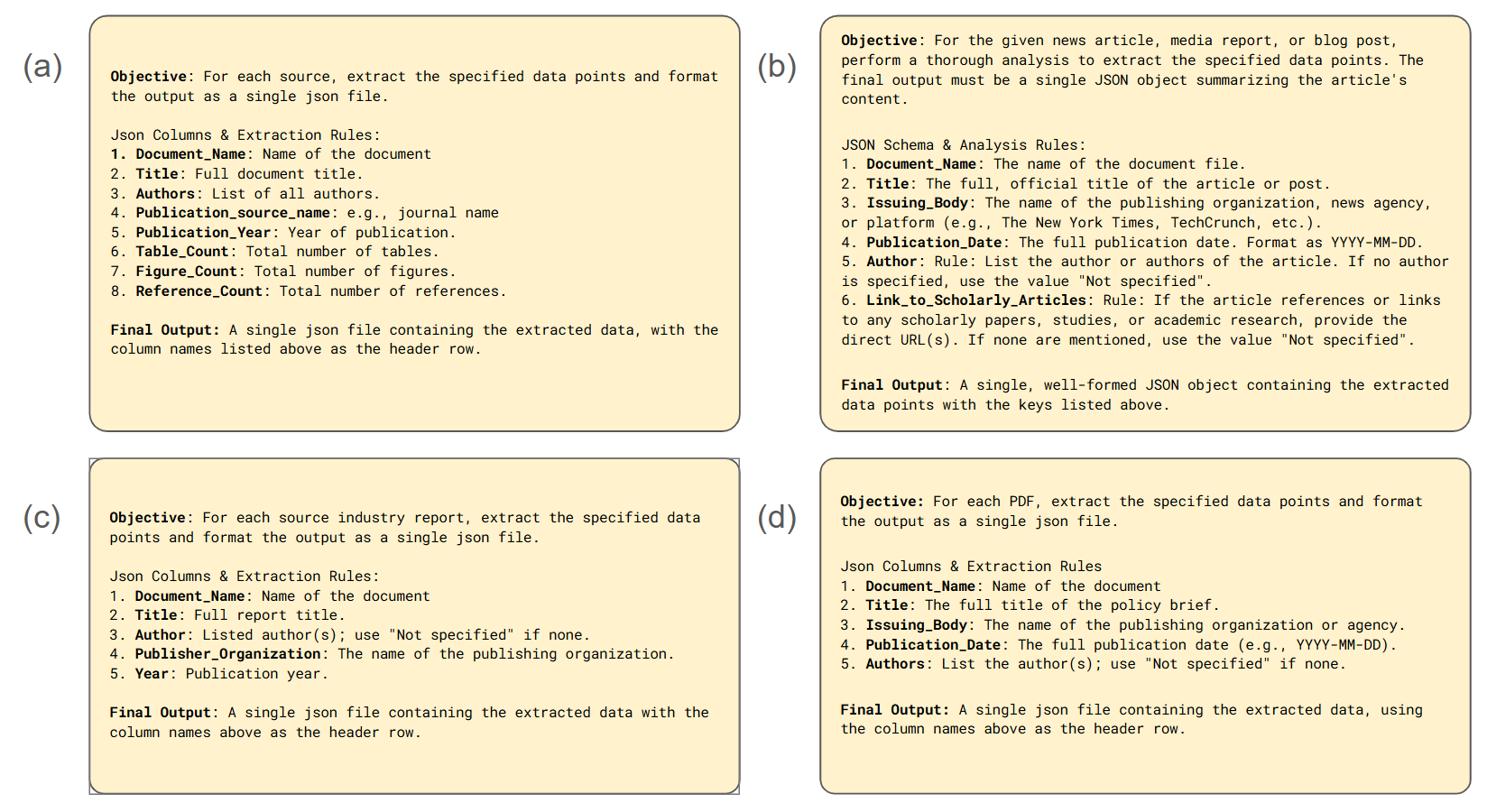}
    \caption{\textbf{Extraction prompts accross source types.}
\textbf{(a) Research papers:} extraction prompt designed to capture bibliographic and structural metadata, including title, authors, journal/source name, publication year, and counts of tables, figures, and references. \textbf{(b) News, media and blogs:} schema records outlet or issuing body, publication date, author(s), title, and links to cited research when available. \textbf{(c) Industry reports:} schema emphasizes publishing organization, authorship, year, title, and report type, reflecting organizational rather than academic provenance. \textbf{(d) Policy briefs:} schema adapted for policy sources, extracting issuing body, publication date, authorship, and title. schema records issuing body, publication date, author(s), and title. Across all categories, the output is a single JSON object per document to ensure consistent, machine-readable metadata for downstream analysis.}
    \label{fig:fig5}
\end{figure}

\begin{figure}[H]
    \includegraphics[width=1\textwidth]{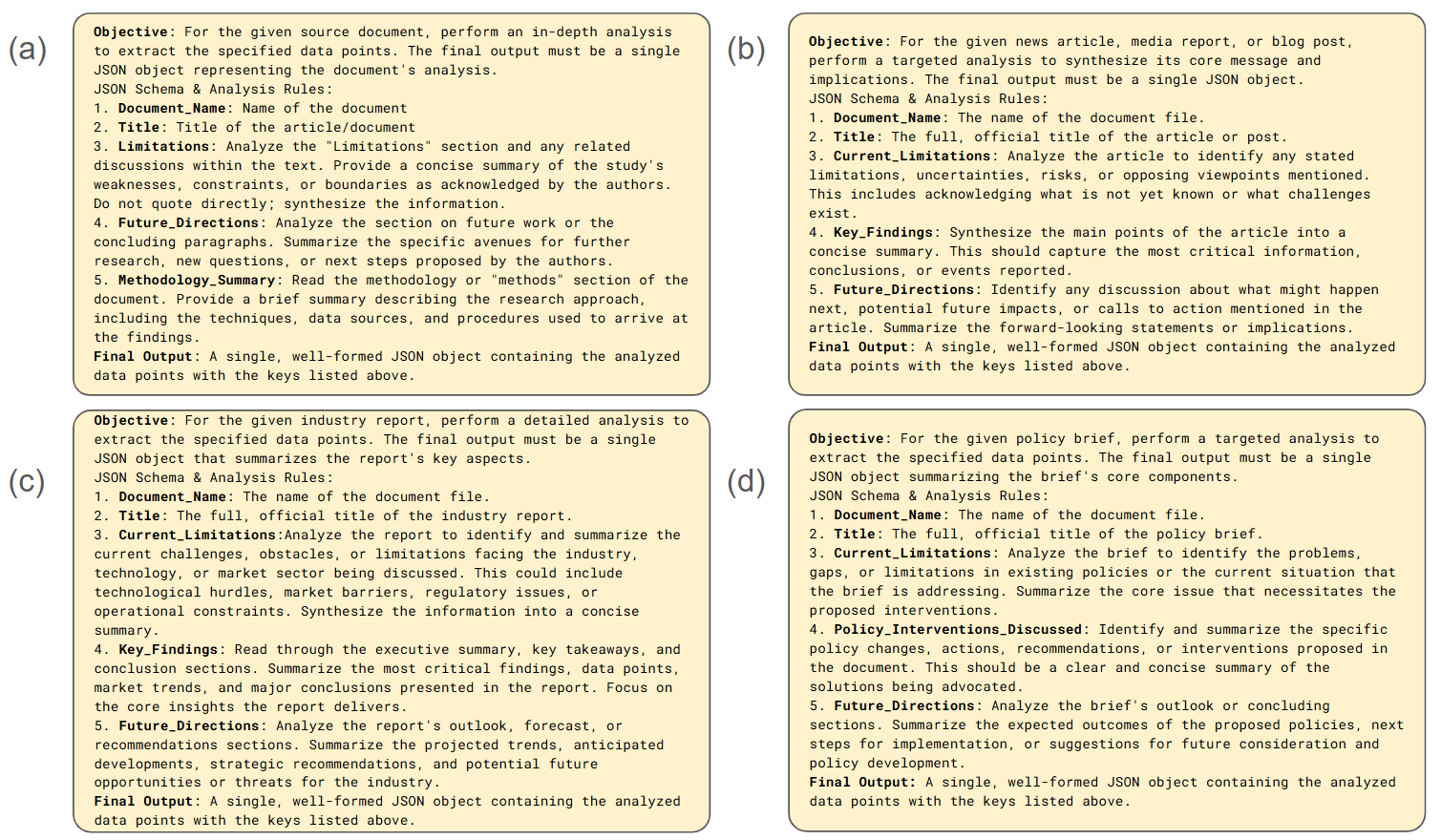}
    \caption{\textbf{Analysis prompts across source types
(a) Research papers:} extract and synthesize study limitations, future research directions, and a concise methodology summary. \textbf{(b) News, media, and blogs:} identify current limitations, key findings, and forward-looking implications or calls to action. \textbf{(c) Industry reports:} capture current constraints (technological, market, regulatory), core market findings, and projections or strategic recommendations. \textbf{(d) Policy briefs:} surface policy problems or gaps, proposed interventions, and future directions for development and evaluation. All outputs are standardized as well formed JSON objects to enable consistent comparison across sources in downstream analyses.}
    \label{fig:fig6}
\end{figure}

\begin{figure}[H]
    \centering
    \includegraphics[width=1\textwidth]{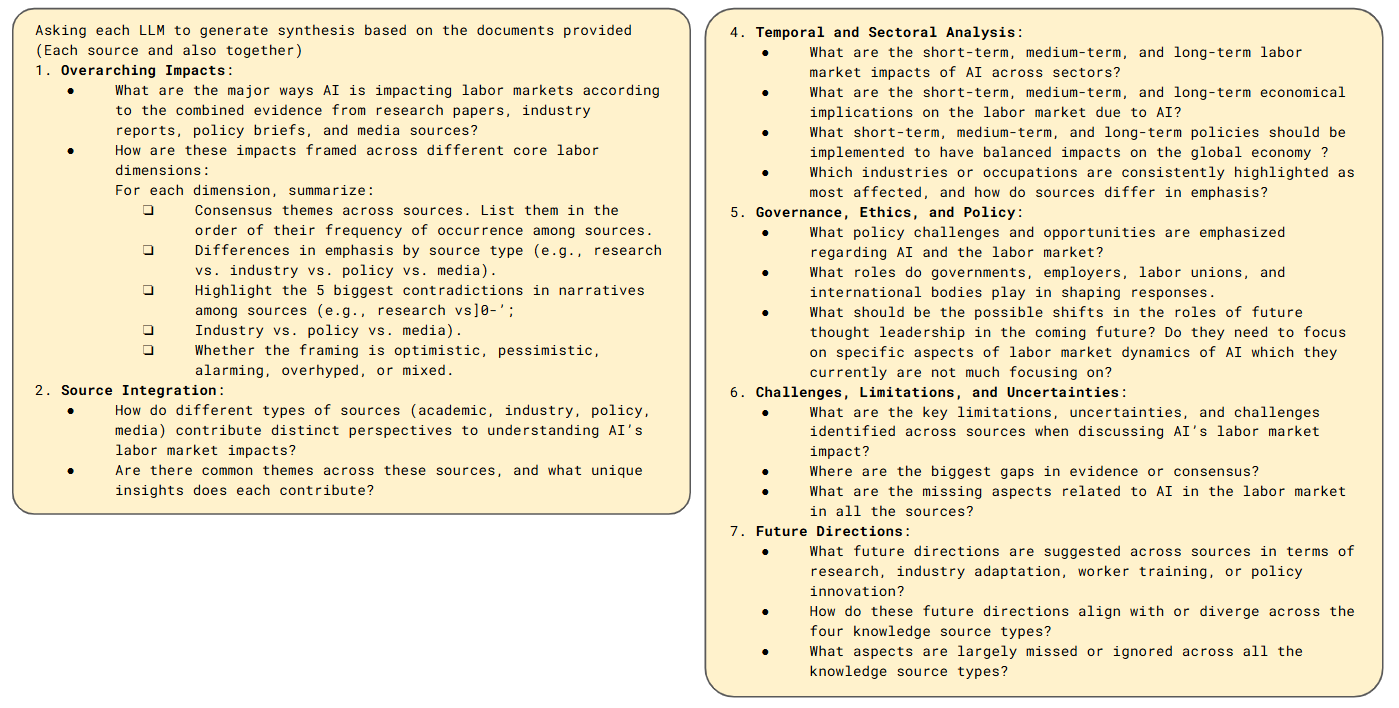}
    \caption{\textbf{Synthesis prompt across source types.}
The synthesis prompt directs LLMs to integrate findings from research papers (RPs), industry reports (IRs), policy briefs (BPs), and news, media, and blogs (NMBs) by addressing six dimensions:\textbf{(1) Overarching impacts:} major ways AI affects labor markets, noting areas of consensus, divergence, contradiction, and framing; \textbf{(2) Source integration:} comparison of contributions and perspectives across RPs, IR, PBs, and NMBs; \textbf{(3) Temporal and sectoral analysis:} hort-, medium-, and long-term effects across sectors, occupations, and economic outcomes; \textbf{(4) Governance, ethics, and policy:} analyzing regulatory challenges, stakeholder roles, and leadership priorities; \textbf{(5) Challenges, limitations, and uncertainties:} evidence gaps, risks, and unresolved debates; and \textbf{(6) Future directions:} proposed research, industry practices, workforce training, and policy priorities. Outputs are captured as a structured summary to support consistent cross-source comparison.}
    \label{fig:fig7}
\end{figure}
\end{appendices}

\end{document}